\documentstyle[12pt,epsfig]{article}
\newcommand{\ba}{\begin{array}}
\newcommand{\ea}{\end{array}}
\newcommand{\beq}{\begin{equation}}
\newcommand{\eeq}{\end{equation}}
\newcommand{\bea}{\begin{eqnarray}}
\newcommand{\eea}{\end{eqnarray}}

\def\bce{\begin{center}}
\def\ece{\end{center}}

\def\nonu{\nonumber}

\def\pa{\partial}

\def\be{\beta}

\def\de{\delta}

\def\ep{\epsilon}

\def\la{\lambda}
\def\La{\Lambda}

\def\si{\sigma}

\def\eps6{{\displaystyle \mathop{\epsilon}^{6}}{}}

\def\nab6{{\displaystyle \mathop{\nabla}^{6}}{}}

\def\ba{\begin{array}}
\def\ea{\end{array}}
\def\beq{\begin{equation}}
\def\eeq{\end{equation}}
\def\be{\begin{equation}}
\def\ee{\end{equation}}

\def\d{\partial}
\def\la{\lambda}
\def\eps{\epsilon}

\def\d{{\rm d}}

\def\ba{\begin{array}}
\def\ea{\end{array}}
\def\beq{\begin{equation}}
\def\eeq{\end{equation}}
\def\be{\begin{equation}}
\def\ee{\end{equation}}

\def\d{\partial}
\def\la{\lambda}
\def\eps{\epsilon}

\def\d{{\rm d}}

\newcommand{\bean}{\begin{eqnarray*}}
\newcommand{\eean}{\end{eqnarray*}}

\begin{document}
\thispagestyle{empty} \addtocounter{page}{-1}
\begin{flushright}
{\tt hep-th/0412202}\\
\end{flushright}

\vspace*{1.3cm} 
\centerline{ \Large \bf ${\cal N}=2$ Conformal Supergravity 
from Twistor-String Theory }
\vspace*{1.5cm}
\centerline{{\bf Changhyun Ahn
}} 
\vspace*{1.0cm} 
 \centerline{\it  School of Natural Sciences,
Institute for Advanced Study,
Einstein Drive, Princeton NJ 08540, USA}
\centerline{\it Department of Physics,
Kyungpook National University, Taegu 702-701, Korea}
\vspace*{0.8cm} 
\centerline{\tt
ahn@ias.edu}  
\vskip2cm

\centerline{\bf Abstract}
\vspace*{0.5cm}

A chiral superfield strength in 
${\cal N}=2$ conformal supergravity at linearized level is obtained  
by acting two superspace derivatives on ${\cal N}=4$ chiral 
superfield strength which can be described in terms of  
${\cal N}=4$ twistor superfields.    
By decomposing $SU(4)_R$ representation  of
${\cal N}=4$ twistor superfields into the  
$SU(2)_R$ representation with  
an invariant $U(1)_R$ charge, 
the surviving ${\cal N}=2$ twistor superfields 
contain the physical states of 
${\cal N}=2$ conformal supergravity. 
These ${\cal N}=2$ twistor superfields
are functions of homogeneous coordinates of weighted complex 
projective 
space ${\bf WCP}^{3|4}$ where the two weighted 
fermionic coordinates have 
weight $-1$ and 3.

\baselineskip=18pt
\newpage
\renewcommand{\theequation}
{\arabic{section}\mbox{.}\arabic{equation}}

\section{Introduction and Summary}
\setcounter{equation}{0}

\indent

There exist two descriptions for twistor-string vertex operators,
topological B-model of
${\bf CP}^{3|4}$ \cite{Witten} and 
open string version of twistor-string theory \cite{Berk}.
A twistor space function of definite {\it homogeneity} describes
a massless particle state in Minkowski spacetime of 
definite {\it helicity} \cite{Pen,Ati}.
The twistor space fields describing ${\cal N}=4$ conformal
supergravity  are homogeneous function of  
bosonic and fermionic variables of ${\bf CP}^{3|4}$ \cite{BW}.
The vertex operators describe supermultiplets 
whose bottom component is independent of fermionic variables 
$\psi^A$ where $A=1,2,3,4$ 
and top component is quartic in $\psi^A$ of ${\bf CP}^{3|4}$. 
Then one can determine $SU(4)_R$ representation for 
each component field in $\psi^A$-expansion by tensor product 
since $\psi^A$ transforms $\bf 4$ under 
that representation.

By looking at the $SU(4)_R$ representations for 
component fields, each supermultiplet
possesses 
8 bosonic and $-8$ fermionic degrees of freedom.
As a result, four 
Lorentz scalar twistor functions have the maximal helicity
$2, 2, 0, 0$ and the minimal helicity $0, 0, -2, -2$.
For the remaining 
eight fermionic twistor supermultiplets, 
the maximal helicity is given by
$3/2$ and $1/2$ and the minimal helicity is $-1/2$ and $-3/2$. 
Since there
are 12 supermultiplets, 
96 bosonic and $-96$ fermionic degrees of freedom are present.
This indicates that the on-shell degrees of freedom in 
${\cal N}=4$ conformal supergravity should reflect this 
observation.

The basic variable of ${\cal N}=4$ linearized conformal 
supergravity \cite{Siegel,BDD,FT} 
is characterized by a chiral superfield strength 
which is a Lorentz scalar.
One of the lessons in \cite{BW} is that they interpreted this 
quantity as a coupling of the abelian gauge field, relevant to
one of the physical vertex operators,  to the 
boundary of an open string world sheet \cite{Berk} in ${\cal N}=4$
superspace. 
The fermionic coordinates $\theta^{A}_{a}$ where $a$ is a 
spinor index 
in this chiral Minkowski superspace
are connected by above fermionic coordinates $\psi^A$ of
${\bf CP}^{3|4}$ via
twistor equations. 
By plugging the solutions for the equations of motion on 
various conformal supergravity fields into a chiral superfield
strength which has all the component fields in 
$\theta_a^A$-expansion, it was 
possible to write the 12 twistor supermultiplets in terms of 
physical states of ${\cal N}=4$ conformal supergravity. 

What happens for ${\cal N}=1$ conformal supergravity \cite{FZ}?
In this case, since 
a chiral superfield strength has three spinor 
indices, a coupling of 
the abelian gauge field to the boundary 
of an open string world sheet should be modified by multiplying
a triple product of ${\cal N}=4$ super derivatives, contracted 
with $SU(4)_R$ indices, which has also three spinor indices.  
The twistor space fields describing ${\cal N}=1$ conformal
supergravity  are homogeneous function of  
bosonic and fermionic variables of 
${\bf WCP}^{3|2}(1,1,1,1|1,3)$ 
\cite{Ahnsept}. See also \cite{PW}.
Four Lorentz scalar functions have the maximal helicity
$2, 2, 0, 0$ and the minimal helicity $3/2, 3/2, -1/2, -1/2$.
For the  
two fermionic supermultiplets, the maximal helicity is
given by 
$3/2$ and $1/2$ and the minimal helicity is given by 
$-1/2$ and $-3/2$. 

The presence of a {\it weighted} fermionic coordinate of weight 3 of
${\bf WCP}^{3|2}$ in 
these fermionic supermultiplets which have an unweighted 
fermionic index   
was crucial to match with the full structure of the physical 
states for ${\cal N}=1$
conformal supergravity.   
As a result, each bosonic supermultiplet
possesses 1 bosonic and $-1$ fermionic degrees of freedom. 
Each fermionic supermultiplet
possesses 2 bosonic and $-2$ fermionic degrees of freedom. 
Therefore, there exist
8 bosonic and $-8$ fermionic degrees of freedom by summing up each 
bosonic and fermionic degree of freedom.
This is consistent with the number of 
on-shell degrees of freedom in 
${\cal N}=1$ conformal supergravity.

The immediate question is to ask how the physical states of
${\cal N}=2$
conformal supergravity \cite{BDD,FT} occur in twistor-string theory.
In the ${\cal N}=2$ conformal supergravity side, 
the number of spinor index of chiral superfield strength is equal to
2 with two $SU(2)$ internal indices that are antisymmetric.
Then  
a coupling of 
the abelian gauge field to the boundary 
in the twistor-string theory should be changed by multiplying
{\it two} ${\cal N}=4$ super derivatives, contracted 
with $SU(4)_R$ indices.
Four Lorentz scalar twistor functions will have the maximal helicity
$2, 2, -1, -1$ and the minimal helicity $1, 1, -2, -2$.
For the  
three fermionic twistor 
supermultiplets, the maximal helicity will be
$3/2$ and 0 and the minimal helicity will be 0 and $-3/2$. 
Each bosonic and fermionic supermultiplet
has 4 bosonic and $-4$ fermionic degrees of freedom. 
There exist
20 bosonic and $-20$ fermionic degrees of freedom which will be
consistent with the number of 
on-shell degrees of freedom in 
${\cal N}=2$ conformal supergravity.

Then what is corresponding twistor-string theory which will contain 
the physical states of ${\cal N}=2$ conformal supergravity? 
As suggested in \cite{Ahnsept}, 
the Calabi-Yau supermanifold should contain the bosonic submanifold
${\bf CP}^3$ of ${\bf CP}^{3|4}$ and the fermionic submanifold 
should have two fermionic coordinates of weight 1 which will
participate in ${\cal N}=2$ chiral Minkowski 
superspace. The superconformal algebra $SU(2,2|2)$ 
will act on these two
fermionic coordinates as well as four bosonic ones.
In order to require the Calabi-Yau supermanifold condition, 
the equality of the sum of  
bosonic weights and the sum of fermionic weights, 
the sum of the remaining fermionic weights should be equal to
2. When we consider {\it two odd} fermionic weights, 
the simplest one is given by 3 and $-1$.  

Therefore, one 
can add homogeneous bosonic and fermionic coordinates of 
${\bf WCP}^{3|2}$ we have considered for ${\cal N}=1$
conformal supergravity to two 
fermionic coordinates of weights $1, -1$  
and make a weighted complex projective space
${\bf WCP}^{3|4}(1,1,1,1|1,1,-1,3)$ where 
the two fermionic coordinates are  {\it weighted} and 
their values are
$-1$ and 3 respectively. Then we will see that twistor
space fields corresponding to the physical states of 
${\cal N}=2$ conformal  supergravity 
are functions of homogeneous coordinates of
this Calabi-Yau supermanifold
\footnote{Naively, one can 
think of ${\bf WCP}^{3|3}(1,1,1,1|1,1,2)$ space by adding 
one even fermionic weight 2 as well as two odd
fermionic weights of 1. 
According to \cite{Sae}, this space 
can be obtained from both ${\bf CP}^{3|4}$ space and 
${\bf WCP}^{3|4}$ 
space by the fermionic dimensional
reduction. Although ${\bf WCP}^{3|4}$ space cannot be obtained
from ${\bf CP}^{3|4}$ through the fermionic dimensional 
reduction directly, since ${\bf WCP}^{3|4}$ space is related to
${\bf WCP}^{3|3}$ space, eventually  ${\bf WCP}^{3|4}$ 
is associated with 
${\bf CP}^{3|4}$ space through ${\bf WCP}^{3|3}$ space.
It might be interesting to see how ${\bf WCP}^{3|3}$ Calabi-Yau
supermanifold can provide ${\cal N}=2$ conformal 
supergravity spectrum.}.  

In section 2, 
by an appropriate decomposition of $SU(4)_R$ representation
of ${\cal N}=4$ twistor fields
into $SU(2)_R$ representation of ${\cal N}=2$ twistor fields, 
the complete structure of helicity states with $SU(2)_R$ 
representation is given.
In section 3,  
by studying the ${\cal N}=2$ linearized conformal 
supergravity, 
the spectrum of massless helicity states
is identified with the twistor-string theory 
description given in section 2.
In section 4, we will make some comments on 
the case of ${\cal N}=3$ conformal supergravity from 
twistor-string theory and a mirror symmetry of 
the above Calabi-Yau supermanifold.

\section{Vertex operators and 
spectrum of massless fields in Minkowski spacetime }
\setcounter{equation}{0}
\label{spectrum}

\indent

In the open twistorial string theory \cite{Berk}, the worldsheet
action depends on the homogeneous coordinates 
\bea
Z^I =(\la^a, \mu^{\dot{a}}, \psi^A)
\nonu
\eea 
of
${\bf WCP}^{3|4}$ and conjugate super twistor variables $Y_I$.
The physical states are given by conformal dimension 1 vertex
operator.
For example, the conformal supergravity multiplet can be  
described by this dimension 1 vertex operator: $Y_I f^I(Z)$ and 
$g_I(Z) \pa Z^I$.
For the multiplets $f^I$ of 
weight 1, they transform as $ f^I \rightarrow t f^I$
under the transformation $Z^I \rightarrow t Z^I$
where $t$ is real. In other words, these $f^I$ carry 
$GL(1)$ charge 1.
For the fermionic multiplets of weight $-1, 3$, 
they scale as $f^{A=3} \rightarrow t^{-1} f^{A=3}$ and 
$f^{A=4} \rightarrow t^3 
f^{A=4}$ under the above transformation 
respectively. They have $GL(1)$ charge $-1$ and 3. 
Similarly, the multiplets $g_I$ of weight $-1$, they transform as 
$ g_I \rightarrow  t^{-1} g_I$ and other fermionic multiplets 
scale as 
$g_{A=3} \rightarrow t g_{A=3}$ and $g_{A=4} \rightarrow t^{-3}
g_{A=4}$
respectively. 
The $GL(1)$ charge of $g_I$ is opposite to those of
$f^I$.

Let us first consider  the twistor field $f^I(Z)$
which is a function of four bosonic and four fermionic variables 
$\la^a, \mu^{\dot{a}}$ and $\psi^A$ \footnote{In this section, the 
twistor fields $f^I(Z)$ an $g_I(Z)$ are ${\cal N}=2$
contents. Unfortunately, later, 
we also denote them by ${\cal N}=4$ twistor fields. Their relations
will
be clear when we discuss about the identifications with 
twistor fields 
in section \ref{identification}.  }. 
Let us denote half of fermionic variables by $\psi$ and $\chi$
which will participate in the ${\cal N}=2$ chiral Minkowski 
superspace, through
the twistor equation,
and the remaining variables by $\alpha$ and $\beta$ which are 
weighted
differently: 
\bea
\psi^{A=1} \equiv \psi, \qquad \psi^{A=2} \equiv \chi; 
\qquad \psi^{A=3} \equiv \alpha, \qquad \psi^{A=4} \equiv 
\beta. 
\nonu
\eea
Since 
the $f^I(Z)$, for each $I=a, \dot{a}, A=1, A=2$, 
is homogeneous in $Z^I$
of degree 1, there exist four bosonic and two fermionic helicity
states, each of helicity $3/2$. Here a massless state in Minkowski 
spacetime has a helicity $1 +\frac{\mbox{deg.}}{2}$ \cite{Pen,Ati} 
when
we put $\psi^A=0$ and do not take the spinor index.
For $A=3$, the $f^{A=3}(Z)$ is homogeneous 
in $Z^I$ of degree $-1$, there is one fermionic helicity 
state $1/2$, by above formula. For $A=4$,
the $f^{A=4}(Z)$ is homogeneous 
in $Z^I$ of degree $3$, there is one fermionic helicity 
state $5/2$ which is greater than 2.
Both spinor index $a$ and $\dot{a}$ provides 
a helicity $1/2$ and $-1/2$. Then this intermediate consideration
leads to two bosonic 
states of helicity 2 and two of helicity 1. 

For the fermions, the index $A=1,2$ transforms
as $\bf 2$ of the $SU(2)_R$ group of R-symmetries and there exist
two states of helicity $3/2$, one state of helicity $1/2$ and 
one state of helicity $5/2$ respectively. 
According to the arguments of 
\cite{BW}, after taking account of the gauge invariance and the 
constraint, there exist two bosonic states of helicity 2 and two
fermionic states  of helicity $3/2$, one fermionic state of helicity 
$1/2$ and one fermionic state of helicity $5/2$.  

The action of 
$SU(2,2|2)$ on the coordinates
$(\la^a, \mu^{\dot{a}}, \psi, \chi)$ of ${\bf WCP}^{3|4}$
is generated by $6 \times 6$ supertraceless matrices where the 
trace of $4\times 4$ upper-left part is equal to 
the trace of $2 \times 2$ lower-right part. 
The bosonic conformal algebra $SU(2,2)$ is represented by
the 15 matrices which lie in the $4 \times 4$
upper-left part. 
A bosonic 
generator which takes the form 
 $\psi^A\frac{\pa}{\pa \psi^A}$ where $A=1,2$ 
for the chiral $U(1)_R$ transformation (which 
does not  exist for ${\cal N}=4$ conformal supergravity) 
is represented by a diagonal supertraceless matrix. 
The 8 spinorial and 8 special conformal symmetry generators
which have the following form $\la^a \frac{\pa}{\pa \psi^A}$,
$\psi^A \frac{\pa }{\pa \mu^{\dot{a}}}$, $\psi^A 
\frac{\pa}{\pa \la^a}$, and $\mu^{\dot{a}} 
\frac{\pa}{\pa \psi^A}$ where $A=1,2$ 
are represented by $2\times 4$ lower-left part and
$4\times 2$ upper-right part. Moreover three $SU(2)$ generators
which are necessary to close the anticommutators between
the spinorial generators and special conformal generators 
are represented by  $2 \times 2$  lower-right part.   

Now for nonzero $\psi^A$, one can expand $f^I(Z)$ in powers of
$\psi^A$:
$
f^I(\la,\mu,\psi)=
f_0^I(\la,\mu) +f_{1A}^{I}(\la,\mu) \psi^A +
f_{2AB}^{I}(\la,\mu) \psi^A \psi^B + \cdots
$
where $f^I_k$ is homogeneous
in $\la, \mu$ 
with degree $(1-k)$ and provides a massless state
of helicity $(3/2-k/2)$ 
when we ignore the angular momentum carried 
by the index $I$. 
One can read off each helicity state and $SU(2)_R$
representation  as one considers for nonzero
$\psi^A$.
As we have observed, there exist
one fermionic state of helicity 
$1/2$ characterized by the twistor field $f^{A=3}(Z)$.
In order to make comparison with ${\cal N}=4$ twistor field
contents, it is better to add a helicity $1/2$ by 
differentiating $f^{A=3}(Z)$ with respect to the third
fermionic coordinate $\psi^{A=3} =\alpha$ and making 
a twistor field which is homogeneous of degree 0 resulting
in a massless state of helicity 1.
  
We list the full structure 
of helicity states described by the ${\cal N}=2$ field $f^I(Z)$
by taking account of angular momentum, the gauge invariance and
the constraint
\bea
\la^a f_a &:& (2, {\bf 1}), \;\; (\frac{3}{2}, {\bf 2}), \;\;
(1, {\bf 1}), \nonu \\
\mu^{\dot{a}} f_{\dot{a}} &:& (2, {\bf 1}), \;\; (\frac{3}{2}, 
{\bf 2}), 
\;\; (1, {\bf 1}), \nonu \\
f^{A=1,2} &:& (\frac{3}{2}, {\bf 2}), \;\; 
(1, {\bf 3} \oplus {\bf 1}), \;\; (\frac{1}{2}, 
{\bf 2} ), 
\nonu \\
\pa_{\alpha} \; f^{A=3} &:& (1, {\bf 1}), \;\; 
(\frac{1}{2}, {\bf 2}), 
\;\; (0, {\bf 1}), 
\label{ffunction}
\eea
where the first element is the helicity and the second element
is the $SU(2)_R$-representation of $R$-symmetries and 
$ \pa_{\alpha} =\frac{\pa}{\pa \psi^{A=3}} =
\frac{\pa}{\pa \alpha}$.
Let us explain what we obtained in detail.
The field contents in each $f^{I}(Z)$ are 
exactly the truncation of ${\cal N}=4$ twistor fields
satisfying $SU(2)_R$ symmetry with same $U(1)_R$ charge for 
${\cal N}=4$ twistor fields under the 
$SU(4)_R \rightarrow SU(2) \times SU(2)_R \times U(1)_R$.
For example, the helicity state $(\frac{3}{2}, 
\overline{{\bf 4}})_{-1}$ with $U(1)_R$ charge $-1$ 
of ${\cal N}=4$ twistor field $\la^a f_a(Z)$
breaks into $(\frac{3}{2}, ({\bf 1,2 })_{-1}) + 
(\frac{3}{2}, ({\bf 2,1 })_{1})$ where the second element has the 
following representation $(SU(2), SU(2)_R)_{U(1)_R}$.
Then the only state $(\frac{3}{2}, ({\bf 1,2 })_{-1})$ preserving
$U(1)_R$ charge survives.
This is denoted by $(\frac{3}{2},{\bf 2})$ above in a simplified 
notation.
One can analyze other helicity states from ${\cal N}=4$
description and how to break into ${\cal N}=2$ field contents with
correct quantum numbers.

In particular, the ${\cal N}=4$ fields $f^{A}(Z)$ contain
two different $SU(4)_R$ representations for each helicity:
$(1,{\bf 15} \oplus {\bf 1}),(\frac{1}{2},\overline{\bf 20 } \oplus
\overline{\bf 4 }),(0,{\bf 10} \oplus {\bf 6})$.
In other words, they are divided into two parts.
Higher dimensional representations ${\bf 15}$ and 
$\overline{\bf 20 }$ 
which have nonnegative Weyl weights \footnote{For a generic field 
$\Phi_{\mu_1 \mu_2 \cdots \mu_n}$, the flat space action contains 
$\int d^4 x \Phi_{\mu_1 \cdots \mu_n} \pa_{\la_1} 
\cdots \pa_{\la_{2p}}
\Phi_{\nu_1
\cdots \nu_n} + \cdots$. By assumption \cite{FT}, 
the canonical mass dimension
of $\Phi$ is equal to $2-p$. If $\Phi$ does not carry world indices
$(n=0)$, then its Weyl weight $w$ is equal to its canonical weight.
If $n\neq 0$, a new field $\Phi_{\mu_1 \cdots \mu_n} e^{\mu_1}_{a_1} 
\cdots e^{\mu_n}_{a_n}$ is defined with $\widetilde{n}=0$ and 
$\widetilde{w}=2-p$. In this case $w=2-p-n$ where $w(e^{\la}_{a})=1$.
For example, the $SU(4)_R$ representation $\bf 15$, 
$V_{\mu A}^{B}$, has a canonical dimension 1 because it has second
order derivatives and its Weyl
weight
is equal to $1-n=1-1=0$. For the representation $\overline{\bf 20}$, 
$\overline{\xi}_{[AB]}^{\dot{a}C}$, its Weyl weight is equal to
its canonical weight $3/2$(it has first order derivative) 
because it does not have 
world indices.   } 
enter into $f^{A=1,2}(Z)$ of
${\cal N}=2$ twistor fields through the breaking  
$SU(4)_R \rightarrow 
SU(2) \times SU(2)_R \times U(1)_R$ 
while lower dimensional representations ${\bf 1}$, $
\overline{\bf 4 }$ and  ${\bf 6}$ which have negative Weyl weights
\footnote{For the graviton $e_{\mu}^{a}$, 
the canonical weight is equal to 0 because it has fourth order
derivatives
while
its Weyl weight is equal to $0-n=0-1=-1$. For the gravitino 
$\overline{\eta}_{\mu A}^{\dot{a}}$, the Weyl weight is equal to
$-1/2$ since the canonical weight is $1/2$ because it has 
third order derivatives and $n=1$. For 
$\overline{T}_{\mu \nu}^{[AB]}$, the Weyl weight is equal to $-1$
because
its canonical weight is 1 because it has second order derivatives 
and $n=2$.  }
enter into $\partial_{\alpha} f^{A=3}(Z)$ which is homogeneous in 
$Z^I$ of degree $0$ 
resulting in a massless state of helicity 1.       

A zero total number of on-shell(dynamical) degree of freedom 
can be checked by counting each bosonic and fermionic degree of
freedom 
in the spectrum.
Each row in (\ref{ffunction}) has 
bosonic degree of freedom 2 and fermionic degree of freedom $-2$.
There exist five columns. Therefore, there are 10 bosonic 
and $-10$ bosonic degrees of freedom. Their sum is equal to zero.
Also note that the fermionic state of helicity $5/2$ characterized
by $f^{A=4}(Z)$ implies higher spin which is greater than 2.
This cannot be obtained from ${\cal N} =4$ conformal 
supergravity, by same reasons \cite{Ahnsept} 
\footnote{Of course, the twistor-string theory, in general, 
contains more information than conformal supergravity theory 
has. We would like to see how the ${\cal N}=2$ conformal 
supergravity spectrum appears in the context of twistor-string 
theory, as a first step. It is an open problem to study the 
full twistor-string theory without ignoring the state of
helicity $5/2$ and it might lead to an interaction of conformal 
supergravity theory with other matter multiplets for this
extra state. }. 
${\bf WCP}^{3|3}$
Let us consider the twistor field $g_I(Z)$ and the Lorentz scalars
$(\la^a g_a, \mu^{\dot{a}} g_{\dot{a}}, \pa_a g^a, \pa_{\dot{a}} 
g^{\dot{a}})$ are homogeneous of degree $(0,0,-2,-2)$ respectively.
By counting of the gauge invariance and the constraint, 
we are left with two twistor fields of degree 
$-2$ leading to 
two massless states of helicity 0.
The field $g_{A=1,2}$
is homogeneous of weight $-1$ describing 
massless states of helicity $1/2$ while
the field $g_{A=3}$
is homogeneous of weight $1$ describing 
massless states of helicity $3/2$
and the field $g_{A=4}$
is homogeneous of weight $-3$ describing 
massless states of helicity $-1/2$. 
As for the case of $f^{A=4}(Z)$, this $g_{A=4}(Z)$ 
is not allowed
for the spectrum.
As we have observed in $f^I(Z)$, we need to 
act some fermionic differentiation on $g_I(Z)$ in order to
see the truncation of ${\cal N}=4$ twistor fields manifestly.
Then the  
complete structure 
of helicity states 
by the field $g_I(Z)$
are summarized by
\bea
\pa_{\alpha} \pa_{\beta} \; \pa_a  g^a &:& (-1, {\bf 1}), 
\;\; (-\frac{3}{2}, {\bf 2}), \;\;
(-2, {\bf 1}), \nonu \\
\pa_{\alpha} \pa_{\beta} \; \pa_{\dot{a}}  
g^{\dot{a}} &:& (-1, {\bf 1}), \;\; (-\frac{3}{2}, 
{\bf 2}), 
\;\; (-2, {\bf 1}), \nonu \\
\pa_{\alpha} \pa_{\beta} \; g_{A=1,2} &:& (-\frac{1}{2}, {\bf 2}), 
\;\; (-1, {\bf 3} \oplus {\bf 1}), 
\;\; (-\frac{3}{2}, 
 {\bf 2}),
\nonu \\
\pa_{\beta} \; g_{A=3} &:& (0, {\bf 1}), \;\; 
(-\frac{1}{2}, {\bf 2}), 
\;\; (-1, {\bf 1}), 
\label{gfunction}
\eea
where
$ \pa_{\beta} =\frac{\pa}{\pa \psi^{A=4}} =
\frac{\pa}{\pa \beta}$.
The role of $\pa_{\alpha}$ and $\pa_{\beta}$ is to add 
the degree $1,-3$ to
any twistor fields respectively. 
The field contents in each $g_{I}(Z)$ are 
exactly the truncation of ${\cal N}=4$ twistor fields
satisfying $SU(2)_R$ symmetry for the invariant  $U(1)_R$ charge.
In this case, also the ${\cal N}=4$ fields $g_{A}(Z)$ contain
two different representations for each helicity:
$(0,\overline{{\bf 10}} \oplus {\bf 6}),
(-\frac{1}{2},{\bf 20 } \oplus
{\bf 4 }),(-1,{\bf 15} \oplus {\bf 1})$.

Higher dimensional representations ${\bf 20}$ and ${\bf 15 }$ 
of $SU(4)_R$ 
enter into $\pa_{\alpha} \pa_{\beta} g_{A=1,2}(Z)$ of
${\cal N}=2$ twistor fields  
while lower dimensional representations ${\bf 6}$,
${\bf 4 }$, and ${\bf 1}$ 
of $SU(4)_R$
enter into $\pa_{\beta} g_{A=3}(Z)$.   
We will see the detailed analysis in next section.
The massless fields described by (\ref{gfunction})
have the opposite helicities and $U(1)_R$ charges 
from those described by  (\ref{ffunction}).
In this case, there are also 10 bosonic and $-10$ fermionic
degrees of freedom. Therefore, 
a zero total number of on-shell(dynamical) degree of freedom 
is given by 20 bosonic and $-20$ fermionic degrees of freedom
(their sum is equal to zero), 
by counting each bosonic and fermionic degree of
freedom 
in the spectrum of $f^I(Z)$ an $g_I(Z)$.

We will prove that the spacetime fields described by 
the twistor fields $f^I(Z)$ and $g_I(Z)$ 
we have constructed contain the physical states of
${\cal N}=2$ conformal supergravity in next section.

\section{Linearized spectrum of ${\cal N}=2$ conformal supergravity
and twistor description}
\setcounter{equation}{0}
\label{identification}

\indent

We will describe how ${\cal N}=2$ chiral superfield strength arises
from corresponding ${\cal N}=4$ chiral superfield strength,
determine the spectrum of massless helicity states, 
and
compare those with the results obtained from twistor-string 
description.

\subsection{A chiral superfield strength in 
${\cal N}=2$ conformal supergravity}

\indent

The linearized ${\cal N}=4$ conformal supergravity 
can be described by a chiral scalar superfield 
${\cal W}^{{\cal N}=4} (x, \theta)$
whose Weyl weight 0 and whose 
component $\theta$-expansion has the following explicit 
form \cite{BW,BDD}
\bea
{\cal W}^{{\cal N}=4} (x, \theta) &=&
\cdots + 
(\theta^2)^{(AB)} \; E_{(AB)} +(\theta^2)^{(ab)}_{[AB]}
\; T_{(ab)}^{[AB]} \nonu \\
&& +(\theta^3)_D^{(abc)} \; (\pa \eta)_{(abc)}^D+
(\theta^3)_{[AB]}^{aC} \; \xi_{aC}^{[AB]} \nonu \\
&& + 
(\theta^4)_{B}^{A(ab)} \; (\pa V)_{(ab)A}^{B} +
(\theta^4)^{(abcd)} \; W_{abcd} + (\theta^4)_{[CD]}^{[AB]} 
\; d^{[CD]}_{[AB]} \nonu \\
&& + (\theta^5)_{C}^{a[AB]} \; \pa_{a\dot{a}} 
\overline{\xi}_{[AB]}^{\dot{a}C} +(\theta^5)^{A(abc)} 
\; (\pa \rho)_{A(abc)}
 \nonu \\
&& + 
(\theta^6)_{(AB)} \; \pa_{\mu} \pa^{\mu} \overline{E}^{(AB)} 
+ (\theta^6)_{[AB](ab)} \; \pa^{a\dot{a}} \pa^{b\dot{b}} 
\overline{T}_{\dot{a} \dot{b}}^{[AB]} 
 + \cdots
\label{n4w}
\eea
which satisfies
the constraint
\bea
\ep^{ABCD} D^a_C D_{Ea} D_{bD} D^b_F {\cal W}^{{\cal N}=4}
= \ep_{EFGH} \overline{D}^{\dot{a}A} 
\overline{D}_{\dot{a}}^{G} \overline{D}^B_{\dot{b}} 
\overline{D}^{\dot{b}H} \overline{{\cal W}}^{{\cal N}=4}
\label{constraint}
\eea
and the chirality of ${\cal W}^{{\cal N}=4}$ is imposed:
it does not contain the $\overline{\theta}_A^{\dot{a}}$-dependence.
Due to this contraint equation, some of higher $\theta$-components
of ${\cal W}^{{\cal N}=4}$ 
are expressed in terms of derivatives of fields whose
complex conjugates occur in the lower $\theta$-components.
The $SU(4)_R$ irreducible representations 
$(\theta^2)^{(AB)}, (\theta^2)^{(ab)}_{[AB]},
(\theta^3)_D^{(abc)}, \cdots, (\theta^6)_{[AB](ab)}$
transform as ${\bf 10}$, ${\bf 6}$, $\overline{\bf 4}$,
$\overline{\bf 20}$,
${\bf 15}$,${\bf 1}$,
${\bf 20}'$, ${\bf 20}$, ${\bf 4}$, 
$\overline{\bf 10}$, ${\bf 6}$ of $SU(4)_R$ respectively.
In particular, $ \xi_{aC}^{[AB]}$
and $d^{[CD]}_{[AB]}$ are 
traceless. 

Let us construct ${\cal N}=2$ chiral field strength superfield
${\cal W}^{AB}_{ab}(x, \theta)$ where $A$ and $B$ are $SU(2)$ 
indices and antisymmetric 
in terms of above ${\cal W}^{{\cal N}=4}$,
by acting two super derivatives on ${\cal W}^{{\cal N}=4}$
contracted with $SU(4)$-invariant epsilon tensor  
and putting the other
fermionic coordinates 
to zero,
\bea
{\cal W}^{12}_{ab}(x, \theta) = 
\epsilon^{12AB} D_{aA} D_{bB}  {\cal W}^{{\cal N}=4}(x,\theta) 
\Big |_{\theta^3=\theta^4=0}
\label{twoder}
\eea
where we fix $SU(2)$ indices 1 and 2.

We will compute the right hand side of (\ref{twoder}) together with 
(\ref{n4w}) explicitly.
Let us first consider the quadratic term in $\theta$ of 
${\cal W}^{{\cal N}=4}$,
$(\theta^2)^{(CD)}$, 
transforming 
${\bf 10}$ under the $SU(4)_R$ when we act two
super derivatives on it, together with 
$E_{(CD)}$. There are two contributions but they 
are cancelled each other.
That is, one obtains
\bea
\epsilon^{12AB} D_{aA} D_{bB} \; (\theta^2)^{(CD)} \; E_{(CD)}
\Big |_{\theta^3=\theta^4=0}
=0.
\label{quad1}
\eea

Next let us 
compute the contributions from the quadratic term 
in $\theta$ of ${\cal W}^{{\cal N}=4}$,
$(\theta^2)^{(cd)}_{[CD]}$,
transforming $\bf 6 $ under the $SU(4)_R$. 
The result can be written as $\ep^{12EF} \ep_{CDEF} 
\de_{(a}^{c} \de_{b)}^{d}$ by realizing that
$(\theta^2)^{(cd)}_{[CD]}=
\ep_{CDEF} \theta^{E(c} \theta^{d)F}$. 
By combining this with the component field 
$T^{[CD]}_{(cd)}$, one gets 
\bea
\epsilon^{12AB} D_{aA} D_{bB} \; (\theta^2)^{(cd)}_{[CD]}
\; T_{(cd)}^{[CD]}
\Big |_{\theta^3=\theta^4=0}
=
\de_{[C}^{1} \de_{D]}^{2} \; T^{[CD]}_{(ab)}.
\label{quad2}
\eea
Here the role of two super derivatives acting on the 
fermionic coordinates and putting $\theta^3=\theta^4=0$
projects out the representation ${\bf 6}$, 
$T_{(cd)}^{[CD]}$,
of $SU(4)_R$ and leaves 
invariant ${\bf 1}$, $T^{[12]}_{(ab)}$, of $SU(2)_R$.  

Let us compute the contributions from cubic terms 
in $\theta$ of ${\cal W}^{{\cal N}=4}$,
$(\theta^3)_C^{(cde)}$, which has  
the following expression $\ep_{CDEF} \theta^{D(a} \theta^{Eb} 
\theta^{c)F}$,
transforming ${\overline{\bf 4}}$ under the $SU(4)_R$.
It is easy to check   
$\epsilon^{12AB} D_{aA} D_{bB} (\theta^3)_C^{(cde)}=
\de_a^{(c} \de_b^{d} \theta^{e)[1} \de^{2]}_{C}
$ by computing the action of 
super derivatives \footnote{Let us emphasize that
after differentiating the $\theta$'s under the super
derivative $D_{aA}$ and putting the condition $\theta^3=
\theta^4=0$, the
remnants of $\theta^{aA}$'s where $A=1,2$ 
are the fermionic coordinates of ${\cal N}=2$
superspace.}.
Then by combining this with the component field 
$(\pa \eta)_{(cde)}^C$,
one obtains
\bea
\epsilon^{12AB} D_{aA} D_{bB} \; (\theta^3)_D^{(cde)} \; 
(\pa \eta)_{(cde)}^D
\Big |_{\theta^3=\theta^4=0}
=\theta^{c[C} (\pa \eta)^{D]}_{(ab)c} \; \de_{[C}^{1} \de_{D]}^{2}.
\label{cubic1}
\eea
In this case, the only state represented by 
${\bf 2}$, $(\pa \eta)^{D}_{(ab)c}$ where $D=1$ or 2, 
of $SU(2)_R$ is survived by acting  
some projection on ${\bf 4}$ of $SU(4)_R$.
The next term $(\theta^3)_{[CD]}^{cE}=\ep_{CDFG} (\theta^2)^{EF} 
\theta^{Gc}$ 
we consider transforms as $\overline{\bf 20 }$
under the $SU(4)_R$ and the contributions can be written as 
$\ep^{12FG} \ep_{CDFG} 
\de^{c}_{(a} \theta_{b)}^{E}$ and $\ep^{12EF} \ep_{CDFG} 
\de^{c}_{(a} 
\theta_{b)}^G$. The first term combined with 
the component field $ \xi_{cE}^{[CD]}$ becomes 
$ \de_{[C}^{1} \de_{D]}^{2} \theta_{(a}^A \xi_{b)A}^{[CD]}$.
Note that when we restrict the condition of $\theta^3=\theta^4=0$,
the $SU(4)$ index $A$ becomes the $SU(2)$ index. That is, 
the value of index $A$ is 1 or 2. 

What happens for the second
term? The whole expression will be 
$  \ep^{12EF} \ep_{CDFG}
\theta_{(a}^G  \xi_{b)E}^{[CD]}$. 
When the index $G$ is equal to 1 or 2, the index $E$ 
can be $C$ or $D$ in the product of two epsilon tensors. 
However, this contribution vanishes because there exists a relation
$\xi_{bC}^{[CD]}=0$.
Eventually, we end up with 
\bea
\epsilon^{12AB} D_{aA} D_{bB} \; (\theta^3)_{[CD]}^{cE}
 \; \xi_{cE}^{[CD]}
\Big |_{\theta^3=\theta^4=0}=  \theta_{(a}^A \xi_{b)A}^{[CD]}
\; \de_{[C}^{1} \de_{D]}^{2}.
\nonu
\eea
After projecting out, the representation of ${\bf 2}$, 
$ \xi_{bA}^{[12]}$ where $A=1$ or 2, 
of $SU(2)_R$
remains. Originally it was a state represented by
${\bf 20}$ of $SU(4)_R$. 

We move on the quartic term in $\theta$ of ${\cal W}^{{\cal N}=4}$. 
First let us consider 
the term $(\theta^4)_{D}^{C(cd)}$ transforming 
${\bf 15}$ of $SU(4)_R$.
By factorizing this into 
$(\theta^2)^{CE}$ and $(\theta^2)^{cd}_{[DE]}$,
as the two super derivatives act on only the last factor, 
the result can be written as 
$(\theta^2)^{C[D} (\pa V)^{E]}_{C](ab)} \de^{1}_{[D} \de^{2}_{E]}$ 
where $C=1,2$ by using the property of (\ref{quad2}). 
On the other hand, when the two 
super derivatives act on the first and second factors separately,
half of the terms contain $\de_C^D$ and this will lead to
the contribution $(\pa V)^A_A$. This is zero from the 
irreducibility of $SU(4)$ tensor.
The rest of the terms consists of 
$
(\theta^2)_{(a}^{cC[D} 
(\pa V)_{b)cC}^{E]} \de_{[D}^{1} \de_{E]}^{2}$.
Then one obtains
\bea
\epsilon^{12AB} D_{aA} D_{bB} \; (\theta^4)_{D}^{C(cd)}
\; (\pa V)_{(cd)C}^{D}
\Big |_{\theta^3=\theta^4=0} =
(\theta^2)^{A[C} (\pa V)_{(ab)A}^{D]} \; \de_{[C}^{1} \de_{D]}^{2}.
\nonu
\eea
Therefore $(\pa V)_{(ab)A}^{D}$ where the indices $A, D$ are 1 or 2 
transforms as ${\bf 3} + {\bf 1}$
under the $SU(2)_R$.

For the singlet $(\theta^4)^{(cdef)}=
\ep_{CDEF} \theta^{(cC} \theta^{dD} \theta^{eE} \theta^{f)F}$ 
of $SU(4)_R$, 
the contribution can be written as 
\bea
\epsilon^{12AB} D_{aA} D_{bB} \; 
(\theta^4)^{(cdef)} \; W_{cdef}
\Big |_{\theta^3=\theta^4=0}
= (\theta^2)_{cd}^{[CD]} \; W_{(ab)}^{\;\;\;\;cd} \;
\de_{[C}^{1} \de_{D]}^{2}.
\nonu
\eea
%
By rewriting $(\theta^4)_{[EF]}^{[CD]}$ as 
a factorized form $\ep_{EFGH} (\theta^2)^{CG} (\theta^2)^{DH}$,
the only nonzero contribution occurs when each super derivative 
$D$ acts on each $(\theta^2)$ factor because 
we have seen that as
the two super derivatives act on only one factor, 
there is no contribution from (\ref{quad1}).
Among four terms, the only nonzero contribution arises
from the expression $ \de_{[E}^{1} 
\de_{F]}^{2} (\theta^2)_{ab}^{CD}$. 
The other three terms 
vanish when we restrict to the case of $\theta^3=\theta^4=0$
since the $SU(4)$ index $C$ or $D$ should be equal to 
$E$ or $F$ and the irreducibility condition implies 
$d^{[EF]}_{[CE]}=0$. 
Therefore, we are led to  
\bea
\epsilon^{12AB} D_{aA} D_{bB}  \; (\theta^4)_{[EF]}^{[CD]} 
\;
d^{[EF]}_{[CD]}
\Big |_{\theta^3=\theta^4=0}=  
(\theta^2)^{CD}_{(ab)} \; d^{[EF]}_{[CD]}
 \; \de_{[E}^{1} 
\de_{F]}^{2}.
\nonu
\eea
The original representation ${\bf 20}^{\prime}$ of $SU(4)_R$
is reduced to a state by ${\bf 1}$, $ d^{[12]}_{[CD]}$ where 
the indices $C$ and $D$ are 1 or 2,
 of $SU(2)_R$.

Now let us consider the next term 
$(\theta^5)_{E}^{c[CD]}$
of $\bf 20$ under the $SU(4)_R$.
Let us factorize this into $(\theta^2)^{[CD]}_{de}$ and
$(\theta^3)^{cde}_{E}$.
Then there are three possibilities for nontrivial contributions.
When the two super derivatives act on 
the first $(\theta^2)$ factor,
the result turns out to be $\ep^{12CD} (\theta^3)_{E(ab)}^{c}$, 
according to the computations (\ref{quad2}). 
However, after restricting to the condition of $\theta^3=\theta^4=0$,
this will be the product of three $\theta$'s contracted with the 
three indices of epsilon tensor. Then the contribution becomes zero.
The next nontrivial contribution is the case where all the super
derivatives
are acting on the last factor $(\theta^3)$.
Based on the previous computation (\ref{cubic1}), 
the result can be written as 
$(\theta^2)^{[CD]}_{(be} \theta^{e[1}\de^{2]}_{E} \pa_{a)\dot{a}} 
\overline{\xi}_{[CD]}^{\dot{a}E}$. 
In order to simplify this, let us recall that
the quadratic expression of $\theta$ has the following 
terms: $\ep^{CD} (\theta^2)_{be}$ and $\ep_{be} (\theta^2)^{CD}$.
Then by multiplying an extra $\theta$, one obtains
$\ep^{CD} (\theta^3)^{[E}_{(a} \pa_{b)\dot{a}} \overline{\xi}_
{[CD]}^{F]\dot{a}} \de_{[E}^{1} 
\de_{F]}^{2}$.    
Finally, the last contribution arises when 
the each super derivative acts on each factor separately.
Two of six terms contain 
$\ep_{ab}$ and so 
this becomes zero.
Two of them can be written as 
$\de^C_E$ and this can be combined with 
the component field $\overline{\xi}_
{[CD]}^{E}$. However, 
since $\overline{\xi}_{[CD]}^{C}=0$ from the irreducibility of 
$SU(4)_R$, there is no contribution.  
Two of them can be written as 
$\ep^{12CF} \ep_{EFGH} (\theta^2)^{DH} \theta_a^G \de_b^c$.
One can write this as two 
$(\theta^3)^H_a \ep^{DG}$ and 
$(\theta^3)_a^D \ep^{HG}$-dependent terms. 
In both cases, they contain $\ep_{E}^{C}$ which will 
reduce to zero contribution by same reasoning. 
Therefore, we are led to
\bea
\ep^{12AB} D_{aA} D_{bB}  \; (\theta^5)_{E}^{c[CD]}
\; \pa_{c\dot{a}} 
\overline{\xi}_{[CD]}^{\dot{a}E}
\Big |_{\theta^3=\theta^4=0} =
(\theta^3)_{(a}^{[E} \pa_{b)\dot{a}} 
\overline{\xi}_{[CD]}^{F]\dot{a}} \; \ep^{CD}
\de_{[E}^{1} 
\de_{F]}^{2}.
\nonu
\eea

One can factorize the next term $(\theta^5)^{C(def)}$ of 
$\bf 4$ into two factors:
$(\theta^2)^{CD} (\theta^3)_D^{cde}$.
When the two super derivatives act on the first factor,
the contribution will be zero by (\ref{quad1}). 
Let us consider when the two super derivatives act on the 
second factor. According to the computations done before 
(\ref{cubic1}),
one gets $(\theta^2)^{CD} \theta^{cE} \de^F_D \de_{[E}^{1} 
\de_{F]}^{2}
(\pa \rho)_{Cabc}$. 
In order to simplify this, let us recall that
the cubic expression of $\theta$ has the following 
terms and it can be decomposed into: 
$\ep^{CE} (\theta^3)^{cD}$ and $\ep^{DE} (\theta^3)^{cC}$.
As a result, one obtains
$\ep^{12} (\theta^3)^{Cc} (\pa \rho)_{(ab)c C}$.    
Of course, there are other expressions
which can be easily reduced to this.
Moreover, there is an expression when each super derivative 
acts on each factor.
The only nontrivial part can be obtained from the case where 
the $SU(4)$ index 
$A$ is equal to $C$ while the index $B$ is equal to $E,F$ or $G$.   
In this case, also the final result can be written as
the one we wrote above.  
Therefore, we arrive at the final contribution as follows: 
\bea
\epsilon^{12AB} D_{aA} D_{bB} \; (\theta^5)^{C(def)}
\; (\pa \rho)_{C(def)}
\Big |_{\theta^3=\theta^4=0}=
\ep^{12} (\theta^3)^{Cc} \; (\pa \rho)_{(ab)c C}.
\nonu
\eea

By writing down the explicit dependence of $\theta$'s
where 
$(\theta^6)_{(CD)}$ transforms as $\bf \overline{10}$ 
under the $SU(4)_R$,
there exist 30 terms. Six of them have a structure of $\ep_{ab}$
and this will vanish by symmetrizing the indices $a$ and $b$.
Twelve of them contain three product of $\theta$'s contracted with 
$SU(4)$-invariant epsilon tensor and this also
vanishes because at the final step we put $\theta^3=\theta^4=0$.
The remaining terms take the form of 
$\ep^{12IE} \ep_{EFGC} \ep_{HIJD} \theta_{a}^{F} \theta_{b}^{H} 
\theta^{cG} \theta^{J}_{c}$. From this, 
the last three $\theta$'s can be written in terms of 
both $(\theta^3)_{b}^J$ and $(\theta^3)_{b}^G$. 
By multiplying the first factor $\theta_a^F$,
all these expressions have 
$\ep_{ab}$ and there are no contributions at all.  
Therefore, 
we conclude that
\footnote{One can prove this also by writing $(\theta^6)_{(CD)}$ as
 $\ep_{CEFG} \ep_{DHIJ} 
(\theta^2)^{EH} (\theta^2)^{FI} (\theta^2)^{GJ}$. Then the 
nontrivial contribution can be obtained by acting two  
super derivatives on each $(\theta^2)$ factor. In other words, 
we have $\ep^{12AB} \ep_{CEFG} \ep_{DHIJ} 
(\theta^2)^{EH} \left[D_{aA} (\theta^2)^{FI}\right] 
\left[D_{bB} (\theta^2)^{GJ}\right]$. 
After simplifying this, one gets
an expression of $\ep_{CEFG} \ep_{DHIJ} 
(\theta^2)^{EH} (\theta^2)_{(ab)}^{IJ}$. The last factor 
$(\theta^2)_{(ab)}^{IJ}$ contains $\ep_{ab}$ and 
$(\theta^2)_{ab}$. Then the $\ep_{ab}$ does not contribute 
when we make an antisymmetrization and there exists a relation 
$(\theta^2)^{EH} (\theta^2)_{ab} =0$. As a result, the whole
contribution is zero.} 
\bea
\epsilon^{12AB} D_{aA} D_{bB} \; (\theta^6)_{(CD)}
\; \pa_{\mu} \pa^{\mu} \overline{E}^{(CD)} 
\Big |_{\theta^3=\theta^4=0} =0.
\nonu
\eea

Next terms 
can be manipulated similarly.
When we express
$(\theta^6)_{[CD](cd)}$ explicitly, it can be written as 
$(\theta^2)^{EG} 
(\theta^2)^{FH} (\theta^2)_{[GH]cd} \ep_{CDEF}$.
As we did before, let us act the super derivatives on this.
According to previous computation (\ref{quad1}), when 
the two derivatives act on the first two $(\theta^2)$ factors,
the result is zero. 
The first nontrivial contribution can be obtained from
the case where each super derivative acts on the second and third
$(\theta^2)$ factor respectively.
Among four terms, two terms vanish because the remaining 
$\theta$'s contain $SU(4)$ index 3 or 4.
The remaining terms can be written as 
$\de_{[G}^{1} \de_{J]}^{2} \ep_{CDEF} (\theta^2)^{EG}
(\theta^2)^{FJ}_{ad} \ep_{cb}$.
By manipulating the $\theta$'s, 
the quartic term in $\theta$
can be written as $(\theta^4)(\ep^{EF} \ep^{GJ}+ \ep^{EJ} \ep^{GF})
\ep_{ad}$.
Therefore, 
one arrives at $(\theta^4) \pa_{\dot{a}(a} \pa_{b)\dot{b}} 
\overline{T}^{(\dot{a} \dot{b}) [CD]} 
\ep_{CD} \ep^{12}$. 
The second nontrivial contribution arises 
 from the case where the two super derivatives act on only the last 
$(\theta^2)$ factor.
The result turned out to be
$\de_{[G}^{1} \de_{H]}^{2} \ep_{CDEF} (\theta^2)^{EG}
(\theta^2)^{FH} \ep_{da} \ep_{cb}$ which is equal to the previous
one.
Then, one gets the final expression
\bea
\epsilon^{12AB} D_{aA} D_{bB} \; (\theta^6)_{[CD](cd)}
\; \pa^{c\dot{a}} \pa^{d\dot{b}} 
\overline{T}_{\dot{a} \dot{b}}^{[CD]} 
\Big |_{\theta^3=\theta^4=0}=
(\theta^4)\; \pa_{\dot{a}(a} \pa_{b)\dot{b}} 
\overline{T}^{(\dot{a} \dot{b}) [CD]} 
\; \ep_{CD} \ep^{12}.
\nonu
\eea

After combining all the nonzero contributions we obtained so far,
the component field $\theta$-expansion of ${\cal N}=2$ 
chiral superfield strength ${\cal W}_{ab}^{AB}(x, \theta)$ 
where two $SU(2)_R$ indices $A$ and $B$ are antisymmetric  
is
given by
\bea
{\cal W}_{ab}^{AB}(x,\theta)& = 
& T_{(ab)}^{[AB]} + 
\theta^{c[A} (\pa \eta)_{(ab)c}^{B]}+
\theta^{C}_{(a}  \xi_{b)C}^{[AB]} \nonu \\
&& +
(\theta^2)^{C[A} (\pa V)_{(ab)\;C}^{B]} + 
(\theta^2)^{[AB]}_{cd} F_{(ab)}^{\;\;\;\;\;cd}  
+ (\theta^2)^{[AB]} (\pa A)_{(ab)}+ 
(\theta^2)^{[CD]}_{(ab)} d_{[CD]}^{[AB]} \nonu
\\ && +
(\theta^3)^{[A}_{(a}  \pa_{b)\dot{c}} 
\overline{\xi}^{B]\dot{c}}_{[CD]} \ep^{CD} 
+ (\theta^3)^{cC} (\pa \rho)_{(ab)cC} \ep^{AB} 
\nonu \\ 
&& +
(\theta^4) \ep^{AB} \ep_{CD} \pa_{\dot{a}(a} \pa_{b)\dot{b}} 
\overline{T}^{(\dot{a}\dot{b})[CD]}
\label{Wexpression}
\eea
which was written in \cite{BDD} using four spinor notation
and has Weyl weight $w=1$.
The $SU(2)_R$ irreducible representations 
$ \theta^{cA},
\theta^{C}_{a},
 \cdots, (\theta^4)$
transform as ${\bf 2}$, ${\bf 2}$, ${\bf 3}$,
${\bf 3}$,
${\bf 3}$, ${\bf 3}$,
${\bf 2}$, ${\bf 2}$, ${\bf 1}$, of $SU(2)_R$ respectively.
This is not an unrestricted chiral superfield because
some of higher $\theta$-components  are expressed in terms of
derivatives of fields whose complex conjugates occur in the 
lower $\theta$ sector. One can see this constrained condition 
from ${\cal N}=4$ 
description.
From the constraint equation (\ref{constraint})
in ${\cal N} =4$ superspace,
one sees immediately that 
$D^a_1 D^b_2 {\cal W}_{ab}^{12} = \overline{D}_{\dot{a}}^{1} 
\overline{D}_{\dot{b}}^{2} \overline{{\cal W}}^{\dot{a} 
\dot{b}}_{12}$.
Then the 
linearized ${\cal N}=2$ conformal supergravity can be described
off-shell
by a chiral field strength superfield  
${\cal W}_{ab}^{AB}$ (\ref{Wexpression})
that satisfies the constraint \cite{BDD,FT}
\bea
D^a_A D^b_{B} {\cal W}_{ab}^{AB} = \overline{D}_{\dot{a}}^{A} 
\overline{D}_{\dot{b}}^{B} 
\overline{{\cal W}}^{\dot{a} \dot{b}}_{AB},
\label{CON}
\eea
where
the ${\cal N}=2$ superspace derivatives are given by
$
D^a_A = \frac{\pa}{\pa \theta^{A}_{a}} + 
\overline{\theta}_{A}^{\dot{a}} \pa_{a\dot{a}}, 
\overline{D}^A_{\dot{a}} =\frac{\pa }{\pa 
\overline{\theta}_A^{\dot{a}}}$ where $a,\dot{a},A=1,2$. 
One can find the prepotential $V$ for 
${\cal W}_{ab}^{AB}$ by solving the constraints 
(\ref{CON}) directly or adding the constraints (\ref{CON}) 
in the action together with Lagrange multiplier and eliminating
${\cal W}_{ab}^{AB}$. Then one obtains 
${\cal W}_{ab}^{AB}= \ep^{AB} \overline{D}^4 D^2_{ab} 
V$ \cite{gs}. 

The constraint equation (\ref{CON}) ensures that
the reduced superfield (\ref{Wexpression})
contains $24 + 24$ independent bosonic and fermionic 
components, as does the Weyl multiplet.
Then the condition for 
${\cal W}_{ab}^{AB}$ 
to be chiral, $\overline{D}^{\dot{c}C} 
{\cal W}_{ab}^{AB}(x^{a\dot{a}},\theta^{A}_a,
\overline{\theta}_A^{\dot{a}})=0$
implies that ${\cal W}_{ab}^{AB}$ does not depend on 
$\overline{\theta}_A^{\dot{a}}$.
The field ${\cal W}_{ab}^{AB}$ has an analog
in ${\cal N}=1$ super Yang-Mills theory \cite{superspace,WB}
where, at the linearized 
level, the theory is described by 
a chiral superfield ${\cal W}_a$($ \overline{D}_{\dot{a}} 
{\cal W}_{b}=0$) 
satisfying the additional constraint
equation
$
D^a {\cal W}_{a} = \overline{D}_{\dot{a}}  
\overline{{\cal W}}^{\dot{a}}$.
Some of components in the $\theta$-expansion of ${\cal W}_a$ 
are real. For example, 
this constraint implies that 
the auxiliary field $D^a {\cal W}_{a}$ at 
$\theta=0$ is real.
Other components obey Bianchi identities.
Similarly, one can think of 
the auxiliary fields
$
F_{ab}^{\;\;ab} = \ep^{AB} D_{Aa} D_{Bb} {\cal W}^{ab}
$
at
$\theta=0$ 
and this is real according to (\ref{CON}) and 
the Weyl multiplet
$
F_{abcd} = \ep^{AB} D_{Aa} D_{Bb} {\cal W}_{cd}
$
at
$\theta=0$ 
satisfies the Bianchi identities. 

Since the square of the Weyl multiplet, contracted over
all indices, is a chiral scalar multiplet with Weyl weight
$w=2$. The highest component, a Lorentz scalar, 
has Weyl weight $w=4$. Note that 
the Weyl weights  for $\theta_a^A$ and $D_A^a$ 
are $-1/2$ and $1/2$
respectively.
Then, the action for ${\cal N}=2$ conformal supergravity 
at the linearized level \cite{Siegel,BDD,FT} is given by
\bea
S= \int d^4 x \int d^4 \theta \; {\cal W}^2, \qquad
{\cal W}^2 = \ep_{AB} \ep_{CD} {\cal W}^{abAB}
{\cal W}_{ab}^{CD}.  
\label{action}
\eea
By computing the $\theta$-integrals with correct normalization, 
this action
leads to the following component action \cite{BDD}
\bea
S & = & \int d^4 x  \left[ 4 T_{(ab)}^{[AB]} \pa^{a\dot{a}} 
\pa^{b\dot{b}} 
\overline{T}_{(\dot{a} \dot{b}) [AB]} -\frac{1}{2} 
\xi_{aC}^{[AB]} \pa^{a \dot{a}} 
\overline{\xi}^{C}_{\dot{a}[AB]} + \frac{3}{2} d^2
 +
\frac{1}{2} (\pa V)_{(ab)B}^{A} (\pa V)^{(ab)B}_{A} \right. 
\nonu \\ && \left. -
(\pa \eta)_{(abc)}^{A} (\pa \rho)^{(abc)}_{A} 
+ \frac{1}{4} C_{abcd} C^{abcd}-\frac{1}{8}
(\pa A)_{(ab)} (\pa A)^{(ab)}
 \right].
\label{n2action}
\eea 
Here $d$ is an auxiliary nonpropagating 
pseudo-scalar field.
In Table 1, we present some properties of these
field contents.
The complete generalized component action \cite{BDD} 
for ${\cal N}=2$ conformal supergravity
was found by constructing the non-linear ${\cal N}=2$ Weyl multiplet
and applying the ${\cal N}=2$ density formula.

\subsection{Physical spectrum of ${\cal N}=2$ 
conformal supergravity}

\indent

To find out the spectrum of the theory,
it is necessary to study the solutions 
of equations of motion.
For a single antisymmetric tensor field 
$T_{\mu \nu}^{AB} =T_{[\mu \nu]}^{[AB]}=
(\sigma_{\mu
  \nu})^{(ab)}
T_{(ab)}^{[AB]}$,
the linearized equation of motion 
from above action (\ref{n2action}) 
is given by
$
\pa^{a\dot{a}} 
\pa^{b\dot{b}} T_{(a b)}^{[AB]}=0
$
and the solution becomes, as given in \cite{BW},
\bea
 T_{(ab)}^{[AB]}= \int d^4 k \de(k^2) e^{ik\cdot x} \left[
\pi_a \pi_b \left(T_{-1}^{[AB]}(k) +i \frac{x_0}{k_0}
 T_{-1}^{\prime[AB]}(k) \right) 
+ \pi_{(a} \widetilde{\tau}_{b)} T_0^{[AB]}(k) \right].
\label{Tsolution}
\eea
In this case, the number of on-shell degrees of
freedom is 3 described by three different helicities $-1,-1',0$.
For two spinors $ \xi_{aC}^{[AB]}$,
the equation of motion is given by
$
 \pa^{a \dot{a}} \xi_{aC}^{[AB]} =0
$
and the solution 
describes a massless state of helicity $-1/2$.

For  two gravitinos $\eta_{\mu }^{Aa}$, 
the linearized equation of motion is
$
\pa^{e\dot{b}} 
\pa^{a (\dot{c}} \pa^{\dot{a}) b} \sigma^{\mu}_{b\dot{b}} 
\eta_{\mu a}^A =0
$
and 
the solution is given by \cite{BW}
\bea
\eta_{\mu a}^A = \sigma_{\mu}^{b\dot{b}} \int 
 d^4 k \de(k^2) e^{ik\cdot x} \left[ \pi_a \pi_b \tau_{\dot{b}} 
\left(
\eta_{-\frac{3}{2}}^{A}(k) + i
\frac{x_0}{k_0}\eta_{-\frac{3\prime}{2}}^{A}(k) \right)
+\pi_{(a} \widetilde{\tau}_{b)} 
\tau_{\dot{b}} \eta_{-\frac{1}{2}}^A(k) +
\widetilde{\tau}_a \widetilde{\tau}_b 
\widetilde{\pi}_{\dot{b}} \eta_{\frac{3}{2}}^A(k)  \right]
\nonu
\eea
where the number of on-shell degrees of freedom 
of conformal gravitinos is $-4$ since there exist four 
independent helicity states.
For three $SU(2)$ gauge fields $V_{\mu A}^{B}$,
the equation of motion is obtained 
and the number of on-shell degrees of freedom of the 
gauge vector is equal to 2 by the helicity states $-1$ and 1.

Finally, for graviton 
$e_{\mu a \dot{a}}$, the equation of motion is
$
\pa^{\dot{a} (c} \pa^{d) \dot{b}} \pa^{a(\dot{c}} \pa^{\dot{d})b} 
\sigma_{b\dot{b}}^{\mu} e_{\mu a \dot{a}} =0
$
and the solution \cite{BW} can be written as 
\bea
e_{\mu a \dot{a}} & = & \sigma_{\mu}^{b\dot{b}} \int
d^4 k \de(k^2) e^{ik\cdot x} \left[\pi_a \pi_b \tau_{\dot{a}}
  \tau_{\dot{b}}
\left( e_{-2}(k) +i \frac{x_0}{k_0} e_{-2^{\prime}}(k) \right)+
\pi_{(a} \widetilde{\tau}_{b)} \tau_{\dot{a}} \tau_{\dot{b}} 
e_{-1}(k)
\right. \nonu \\
&& \left. + 
\widetilde{\tau}_a \widetilde{\tau}_b \widetilde{\pi}_{(\dot{a}}
\tau_{\dot{b})}e_1(k) 
 + \widetilde{\tau}_a \widetilde{\tau}_b \widetilde{\pi}_{\dot{a}}
\widetilde{\pi}_{\dot{b}} \left( e_2(k) +i \frac{x_0}{k_0}
  e_{2^{\prime}}(k) 
\right) \right]
\nonu
\eea
where the gauge transformation is used.
The number of on-shell degrees of freedom of the Weyl graviton is 6.
For theories involving higher derivatives,
the counting of states is nontrivial.
The number of on-shell degree of freedom $\nu$ \cite{FT}
from path-integral methods 
which are different from a canonical approach \cite{FZ}
can be found as follows:
\bea
\nu(T_{ab}^{[AB]}) & = &  
\nu(\overline{T}_{\dot{a} \dot{b}}^{[AB]})= 3, \;\;
\nu(\xi_{aC}^{[AB]}) =
\nu(\overline{\xi}^{\dot{a} C}_{[AB]})=-1, \;\;
\nu(\eta_{\mu}^{Aa}) =
\nu(\overline{\eta}_{\mu A}^{\dot{a}})=-4, \;\;
\nonu \\ 
\nu(V_{\mu A}^B) & =& 2, \;\; 
\nu(A_{\mu})  =  2, \;\;
\nu(d) =0, \;\; 
\nu(e_{\mu}^{a\dot{a}})  = 
6.
\nonu
\eea
Then a zero total number of on-shell degree of freedom 
\footnote{The number of off-shell degree of freedom $n$ \cite{FT}
can be found as follows:
$
n(T_{ab}^{[AB]})  =  3, \;
n(\xi_{aC}^{[AB]}) =-2, \;
n(\eta_{\mu}^{Aa}) =-4, \;
n(V_{\mu A}^B) = 3, \;
n(A_{\mu})  =  3, \;
n(d) =1, \; 
n(e_{\mu}^{a\dot{a}})  = 
5$.
Then a zero total number of off-shell degree of freedom 
can be checked by multiplying the above each degree of freedom 
by the number of each type of the field in the spectrum
$
n_{\mbox{total}} = 6 \times 1 -4 \times 2 -8 \times 2 +3 \times 3+
3 \times 1 + 1 \times 1 + 5 \times 1 =+24-24=0$. }
can be checked by multiplying the above each degree of freedom 
by the number of each type of the field in the spectrum
$
\nu_{\mbox{total}} = 6 \times 1 -2 \times 2 -8 \times 2 +2 \times 3+
2 \times 1 + 0 \times 1 + 6 \times 1 =+20-20=0$.

We summarize the various aspects of ${\cal N}=2$
conformal supergravity in Table 1.

\begin{table}
 \begin{tabular}{ccccc} 
  \hline
States in ${\cal N}=2$ CSG & $U(1)_R$ charge & 
Helicity  & $SU(2)_R$ rep. & $SU(4)_R$ rep. 
\\ 
\hline
\vspace{0.3cm}
  $T_{(ab)}^{[AB]}=\ep^{AB} T_{(ab)}$ 
&  $2$ & $-1, -1',0$          &   
${\bf 1}$        & ${\bf 6 }$   
\\
\vspace{0.3cm} 
 $\xi_{aC}^{[AB]}= \ep^{AB} \xi_{aC}$      
& $1$ &  $-\frac{1}{2}$         
&  ${\bf 2}$ & ${\bf 20}$            
\\ 
\vspace{0.3cm}
$\eta_{\mu }^{Aa}$   & $1$ &     $-\frac{3}{2}, -\frac{3}{2}',
-\frac{1}{2}, \frac{3}{2}$      
& ${\bf 2}$ & ${\bf 4}$
   \\ 
\vspace{0.3cm}
$V_{\mu A}^{B}$   & $0$ &     $1,-1$      
& ${\bf 3}$  & ${\bf 15}$           
\\ 
\vspace{0.3cm}
$A_{\mu}$   & $0$ &     $1,-1$      & ${\bf 1}$  & -     
\\
\vspace{0.3cm}
$d_{[AB]}^{[CD]} = \ep_{AB} \ep^{CD} d$   
& $0$      & $\mbox{none}$ &         ${\bf 1}$   & ${\bf 20'}$
\\ 
\vspace{0.3cm}
$e^{a \dot{a}}_{\mu}$   & $0$ &    $2,2',1,-1,-2,-2'$       
&  ${\bf 1}$  & ${\bf 1}$         
\\ 
\vspace{0.3cm}
$\overline{\eta}^{\dot{a}}_{\mu A}$   & $-1$ &
$\frac{3}{2},\frac{3}{2}',\frac{1}{2},-\frac{3}{2}$    
      &  ${\bf 2}$  & ${\overline{\bf 4}}$         
\\ 
\vspace{0.3cm}
  $\overline{\xi}^{\dot{a} C}_{[AB]}=
\ep_{AB} \overline{\xi}^{\dot{a} C}$   & $-1$ &   
$\frac{1}{2}$        &  ${\bf 2}$   & ${\overline{\bf 20}}$        
\\ 
\vspace{0.3cm}
  $\overline{T}_{(\dot{a} \dot{b})}^{[AB]}=
\ep^{AB} \overline{T}_{(\dot{a} \dot{b})}$  & $-2$ 
& $1,1',0$   & ${\bf 1}$     & ${\bf 6}$        
\\ \hline
\end{tabular} 
\begin{center}
\caption{\sl The $U(1)_R$ charges, helicities and $SU(2)_R$
representations of physical states in ${\cal N}=2$ conformal 
supergravity(CSG) in four dimensions. 
In the first column, we present the various 
field contents of ${\cal N}=2$ CSG with $SU(2)_R$-invariant 
epsilon tensor.
The $SU(4)_R$ $R$-symmetry
of ${\cal N}=4$ conformal supergravity is broken to $SU(2) 
\times SU(2)_R \times U(1)_R$. Each $SU(2)_R$ 
representation of
${\cal N}=2$ conformal supergravity corresponds to the 
representation $SU(2) \times SU(2)_R \times U(1)_R$ with invariant
$U(1)_R$ charge. For example, in the branching rule of
${\bf 6} \rightarrow ({\bf 1,1})_{2} +({\bf 1,1})_{-2} +({\bf
  2,2})_{0}$, the $U(1)_R$ invariant representation is 
$({\bf 1,1})_{2}$ which is denoted by $U(1)_R$ charge 2 and 
$SU(2)_R$ representation ${\bf 1}$ separately in the first row.  
All other $SU(4)_R$ 
representations, ${\bf 20, 4, 15, 20',1},\overline{\bf 4}, 
\overline{\bf 20}$ can be broken to the appropriate $SU(2)_R$
representation with invariant $U(1)_R$ charge as above. This 
will be discussed in detail later.}
\end{center}
\end{table}

\subsection{Identification with ${\cal N}=2$ 
twistor fields from ${\cal N}=4$
  twistor fields}

\indent

We will prove that
the physical states of ${\cal N}=2$ conformal 
supergravity in four dimensions 
are equal to the spacetime fields described by the twistor 
fields  $f^I$ and $g_I$.
Let us start with the identification of 
the chiral superfield ${\cal W}_{ab}^{\;\;\;AB}$
in the linearized conformal supergravity with
the twistor fields 
\bea
{\cal W}_{ab}^{12}(x,\theta) = \left( \ep^{12AB} 
D_{aA} D_{bB} \int _{{\bf D}_{x,\theta}} 
g_I dZ^I \right)\Big |_{\theta^3=\theta^4=0},
\nonu 
\eea
where the ${\cal N}=4$ twistor fields $g_I(Z)$ are given in 
\cite{BW} explicitly and the differentials $dZ^I$ are
$ dZ^I =(d \la^a, d \mu^{\dot{a}}, 
d\psi^A) =
(d \la^a, d \la_b x^{b\dot{a}}, d \la_c \theta^{cA})$.
Here we used the twistor equations by introducing a bosonic spinor
$\la_a$
\bea
(\mu^{\dot{a}}, \psi^A)=(x^{a\dot{a}} \la_a, \theta^{Aa} \la_a).
\label{twistor}
\eea

Through these, 
the fermionic coordinates $\psi^A$ of ${\bf CP}^{3|4}$
are related to the fermionic coordinates  
$\theta^A_a$ of chiral Minkowski superspace.
By using the expression of 
$dZ^I$, one can write down
\bea
{\cal W}_{ab}^{12}(x,\theta) = 
 \ep^{12AB} 
D_{A(a} D_{b)B}
\int d \la^c \left[ g_c(Z) +
x_{c\dot{c}} g^{\dot{c}}(Z) + \theta_{c}^{A} g_A(Z) \right]
\Big |_{\theta^3=\theta^4=0}
\label{wab}
\eea
where it is understood that 
the homogeneous coordinates 
$Z^I$ are viewed as functions of $\la^a$
for fixed $x$ and $\theta$ through the twistor equations 
(\ref{twistor}). 
What does this equation (\ref{wab}) mean?
The left hand side of (\ref{wab}) provides 
the solutions for the conformal supergravity fields in momentum
space using the result of previous subsection while 
the right hand side of (\ref{wab}) describes 
the ${\cal N}=2$ twistor fields. 
Our aim is to demonstrate the precise relations
between the physical states of
${\cal N}=2$ conformal supergravity
and 
the vertex operators 
of twistor-string theory using (\ref{wab}).

The equation of motion for ${\cal W}_{ab}^{AB}$
from the superspace action is given by
\bea
D^a_A D^b_{B} {\cal W}_{ab}^{AB}=0. 
\label{equation}
\eea
Let us check that the above superspace function
${\cal W}_{ab}(x,\theta)$ (\ref{wab})
which can be written as twistor fields 
$g_I$ satisfies (\ref{equation}).
Suppose that there is a function $\phi(Z)$ that is 
 any function of
$Z$. Then one can transform it to a function $\phi^{\prime}(\la, x,
\theta)$ by writing $\mu$ and $\psi$ in terms of $\la$
through the relation (\ref{twistor}).
By the chain rule, one gets 
$D_A^a \phi^{\prime}= \la^a \left(\frac{\pa}{\pa \psi^A} + 
\overline{\theta}^{\dot{b}}_{A} \frac{\pa}{\pa \mu^{\dot{b}}} 
\right) \phi^{\prime}$ where $\phi^{\prime}=\phi(\la^c, 
x^{c\dot{c}} \la_c, \theta^{Ad} \la_d)$.  
Since $\la^a \la_a =0$, it is easy to see that 
$D_A^a D_{Ba} \phi^{\prime}=0$.   
As observed in \cite{BW}, inside of an integral (\ref{wab}), there 
are
the factors of $x$ and $\theta$  which are linear in $x$ and 
$\theta$,
the commutators between $D_A^a D_{Ba} $ and $x$ or $\theta$ vanish
and we are left with $D_{A}^{a} D_{Ba} 
D_{C}^{b} D_{Db} 
\int g_I dZ^I=0$ implying that (\ref{wab}) satisfies 
(\ref{equation}). 

Let us determine the ${\cal N}=2$ 
twistor fields  $g_I(Z)$ in terms of 
${\cal N}=2$  chiral superfield ${\cal W}_{ab}^{AB}$ which has
a simple relation (\ref{twoder}) with 
${\cal N}=4$ chiral superfield providing corresponding 
twistor fields.
Let us start with ${\cal N}=4$ twistor field \cite{BW}
\bea
g_{\dot{a}}(Z) 
= \cdots   -\left( i\frac{\la^a \si_{a\dot{a}}^0}{k^0} \right) 
\left[ (\psi^2)_{[AB]} \hat{T}_{-1}^{\prime [AB]}+
 (\psi^3)_A \hat{\eta}^
{\prime A}_{-\frac{3}{2}} + 
(\psi^4) \hat{e}^{\prime}_{-2} \right].
\label{gadot}
\eea
From this, one can take the product of super derivatives 
$D_{c3} D_{d4}$ acting on $g_{\dot{a}}(Z)$
and the nontrivial terms arise from the $\theta^3 \theta^4$-dependent
terms  and
the result is given by \footnote{Since we 
keep the notation for ${\cal
  N}=4$ quantity by $f^I(Z)$ and $g_I(Z)$, the corresponding 
${\cal N}=2$ quantity can be obtained by acting super
derivatives on
${\cal N}=4$ twistor fields and
putting the other fermionic coordinates to zero. 
Therefore, the results depend on only 
$\theta^1$ and $\theta^2$. }
\bea
\epsilon^{12CD} D_{cC} D_{dD}   \; g_{\dot{a}}(Z) 
\Big |_{\theta^3=\theta^4=0}
= (-i\frac{\la^d \si_{d\dot{a}}^0}{k^0} )
 \la_c \la_d   \left[\hat{T}_{-1}^{\prime[AB=12]} +
\psi \hat{\eta}^
{\prime A=2 }_{-\frac{3}{2}} + \chi \hat{\eta}^
{\prime A=1}_{-\frac{3}{2}} + \psi \chi \hat{e}^{\prime}_{-2}   
\right]. 
\label{first} 
\eea
What does this result mean?
We have determined the second term on the right hand side of 
(\ref{wab}) in terms of ${\cal N}=2$ twistor fields.
Then how one can relate this to 
${\cal N}=2$  chiral superfield ${\cal W}_{ab}^{AB}$?
After we are plugging the result (\ref{first}) into the right hand
side of (\ref{wab}), then 
the left hand side of (\ref{wab}) will tell us about 
the component fields for  
${\cal N}=2$  chiral superfield ${\cal W}_{ab}^{AB}$.
In this way, one can determine all the 
identifications between 
the spacetime fields described by twistor fields and 
the physical states of conformal supergravity. 

In section \ref{spectrum},
the helicity and $SU(2)_R$ representation for $\pa_{\alpha} 
\pa_{\beta}
\pa^{\dot{a}}
g_{\dot{a}}$ 
was given in (\ref{gfunction}).
One can identify the right hand side of (\ref{first}) with 
the twistor description in section \ref{spectrum},  term by term. 
For example, the massless state characterized by $(0,{\bf 1})$
before taking $\pa_{\alpha} \pa_{\beta}$ on $\pa^{\dot{a}}
g_{\dot{a}}$  corresponds to 
$\la_c \la_d  \hat{T}_{-1}^{\prime[AB=12]}$ which denotes the twistor
field
for the spacetime field  $  \la_c \la_d 
{T}_{-1}^{\prime[AB=12]}$ of helicity zero \footnote{Recall that 
the hatted fields stand for twistor fields 
and unhatted fields stand for supergravity fields.}
and is a singlet ${\bf 1}$ under
$SU(2)_R$-representation 
\footnote{Let us describe this more explicitly and
how it works.
From the relation
(\ref{wab}), one sees that  $\Box {\cal W}_{cd}= \int d \la^a 
\la_a \ep^{12CD} D_{cD} D_{dD} \pa_{\dot{a}} g^{\dot{a}}$.
Now the lowest term of the left hand side together with 
(\ref{Wexpression}) can be written as
$\int \d^4 k \delta(k^2) e^{i k\cdot x}  \pi_c \pi_d 
{T}_{-1}^{\prime[AB=12]}(k)$ by substituting the solution of equation
of  motion (\ref{Tsolution}) into (\ref{Wexpression}).  
On the other hand, the lowest term of 
right hand side of above relation can be
written as, from the result (\ref{first}),
$\int d \la^a  \la_a \la_c \la_d
\hat{T}_{-1}^{\prime[AB=12]}$  which is equal to $\la_c \la_d 
{T}_{-1}^{\prime[AB=12]}(x)$ up to a numerical factor. This is a 
precise correspondence between the twistor field 
$ \hat{T}_{-1}^{\prime[AB=12]}$
and spacetime 
field ${T}_{-1}^{\prime[AB=12]}$. }. 
The fact that the above spacetime field has a helicity 0 is obvious
because ${T}_{-1}^{\prime[AB=12]}$ itself represents
a helicity $-1$ and  
the role of $\la_c \la_d$ inside the integral of (\ref{wab})
increases the helicity $+1$.
Then how one can understand this $SU(2)_R$ representation?
It is known that 
the $SU(4)_R$-representation ${\bf 6}$
of ${\cal N}=4$ conformal supergravity is decomposed into 
\cite{Slansky}
\bea 
{\bf 6} \rightarrow ({\bf 1,1})_{2} + 
({\bf 1,1})_{-2} +({\bf 2,2})_{0}
\label{sixrep}
\eea
under the 
$SU(4)_R \rightarrow SU(2) \times SU(2)_R \times U(1)_R$.
The only invariant quantity through 
the above procedure(truncation of 
${\cal N}=4$ theory described by (\ref{twoder})) preserving 
$U(1)_R$ symmetry is
$({\bf 1,1})_{2}$ which is represented by 
$\hat{T}_{-1}^{\prime[AB=12]}$ or $T_{-1}^{\prime[AB=12]}$.
Note that when we
take $\pa^{\dot{a}}$ in both sides  of (\ref{first}), the prefactor 
$-i\frac{\la^d \si_{d\dot{a}}^0}{k^0}$ in the right hand side 
is gone \cite{BW}. 
If we are using $\pa_{\alpha} \pa_{\beta}$ on $g_{\dot{a}}(Z)$ 
rather than the super derivatives, then the factor $\la_c \la_d$ 
in the right hand side of (\ref{first}) will disappear. 

What about other terms in the right hand side of (\ref{first})? 
The next massless 
state $(-\frac{1}{2},{\bf
2})$ for $\pa_{\dot{a}} g^{\dot{a}}$ given in section \ref{spectrum}
should correspond to  both 
$ \la_c \la_d \hat{\eta}^
{\prime A=2 }_{-\frac{3}{2}}$ and
$\la_c \la_d \hat{\eta}^
{\prime A=1}_{-\frac{3}{2}}$ in (\ref{first}).
One can combine and write them as $\psi_A 
\hat{\widetilde{\eta}}_{-\frac{1}{2}}^A$
where
index $A$ is a representation ${\bf 2}$ under the 
$SU(2)_R$-representation since the presence of
$\la_c \la_d$ in the spinor integral of (\ref{first}) makes
an increase of helicity $+1$.
The representation ${\bf 2}$ of $SU(2)_R$ can be 
understood as follows. 
More precisely, 
the $SU(4)_R$-representation ${\bf 4}$
is decomposed into
\bea 
{\bf 4} \rightarrow ({\bf 1,2})_{1} + ({\bf 2,1})_{-1}
\label{fourrep}
\eea
under the 
$SU(4)_R \rightarrow SU(2) \times SU(2)_R \times U(1)_R$.
Then the state 
$\hat{\widetilde{\eta}}_{-\frac{1}{2}}^A$ above
corresponds to 
$({\bf 1,2})_{1}$.

Finally, the massless state $(-1,{\bf 1})$ corresponds
to
the last term in (\ref{first}). 
In this case, one can express it as $\psi_A \psi_B
\hat{\widetilde{e}}_{-1}\ep^{AB}$ which has a helicity $-1$
and is a singlet ${\bf 1}$
where the indices $A$ and
$B$ are $SU(2)_R$ indices.

So far, we have 
considered the contributions from $g^{\dot{a}}(Z)$
in (\ref{first}).
Now we move on those from 
$g_a(Z)$.  
Let us consider ${\cal N}=4$ twistor field \cite{BW}
\bea
g_a(Z) = \cdots + \la_a \left[
(\psi^2)_{[AB]} \hat{T}_{-1}^{[AB]}+ (\psi^3)_A 
\hat{\eta}^A_{-\frac{3}{2}} + (\psi^4) \hat{e}_{-2} \right].
\label{ga}
\eea
By acting two super derivatives on this, 
there exists the following relation which looks similar to
(\ref{first}) but the other component fields of `dipole' fields
occur
\bea
\epsilon^{12CD} D_{cC} D_{dD}   \; g_{a}(Z) 
\Big |_{\theta^3=\theta^4=0}
=\la_a \left( \la_c \la_d \right)  \left[  \hat{T}_{-1}^{[AB=12]}+
\psi \hat{\eta}_{-\frac{3}{2}}^{A=2} 
+\chi \hat{\eta}_{-\frac{3}{2}}^{A=1} + 
\psi \chi \hat{e}_{-2}   \right].
\label{seven}
\eea
The factor $\la_a$ in (\ref{seven}) can be removed when 
we consider $\pa^a g_a(Z)$ instead of $g_a(Z)$.
Now it is easy to see the exact identification in
section \ref{spectrum} by following the analysis in previous 
paragraph.
The massless states 
$(0,{\bf 1}),(-\frac{1}{2},{\bf 2}),(-1,{\bf 1})$
given in section \ref{spectrum} 
correspond to the component twistor fields together with 
$\la_c \la_d$ in (\ref{seven}).

Now we are left with the contributions from $\theta^A_c g_A(Z)$
in (\ref{first}).
One has to be careful about these contributions because
the super derivative can act on the fermionic coordinate
$\theta^A_c$ also. 
In this computation, 
let us start with other type of ${\cal N}=4$ twistor field \cite{BW}
\bea
g_{A}(Z) 
& = & \cdots + \psi^B \hat{T}_{0[AB]} + \psi^B \hat{E}_{0(AB)} +
(\psi^2)_{[AB]} \hat{\eta}^B_{-\frac{1}{2}}+
(\psi^2)_{[BC]} \hat{\xi}^{[BC]}_{-\frac{1}{2} A} 
\nonu \\
&& + (\psi^3)_A \hat{e}_{-1} + (\psi^3)_B 
\hat{V}^B_{-1 A} + (\psi^4) \hat{\overline{\eta}}_{-\frac{3}{2} A}.
\label{gA}
\eea
One takes two super derivatives
and has to pick up the 
$\theta^3 \theta^4$-dependent terms in above.
Then it turns out 
\bea
\epsilon^{12CD}  D_{cC} D_{dD}   \; g_{A=1}(Z) 
\Big |_{\theta^3=\theta^4=0}
& = & 
\left( \la_c \la_d \right) \left[
 \left( \hat{\eta}_{-\frac{1}{2}}^{B=2} + 
\hat{\xi}_{-\frac{1}{2},A=1}^{[BC=12]} \right)+
 \psi  \hat{V}_{-1,A=1}^{B=2} \right. \nonu \\
&& \left. +
\chi  \left( \hat{e}_{-1} +  \hat{V}_{-1,A=1}^{B=1} \right) 
 + 
\psi \chi  \hat{\overline{\eta}}_{-\frac{3}{2},A=1}   
\right].
\label{five}
\eea
One has to 
understand this ${\cal N}=2$ twistor field in the context of
(\ref{wab})
and there exists a factor $\theta^{A=1}_{c}$ as well as
(\ref{five}) in the integrand. 
Once again,
after we are plugging the result (\ref{five}) into the right hand
side of (\ref{wab}), then 
the left hand side of (\ref{wab}) will tell us about 
the particular component fields for  
${\cal N}=2$  chiral superfield ${\cal W}_{ab}^{AB}$.
Let us describe each term in detail.
The $SU(4)_R$-representation ${\bf 20}$ represented by 
$\xi_{aC}^{[AB]}$
can be  decomposed into 
\bea
{\bf 20} \rightarrow
({\bf 1,2})_{1} + ({\bf 3,2})_{1} +
({\bf 2,1})_{3} + ({\bf 1,2})_{-3} + ({\bf 2,1})_{-1} +
({\bf 2,3})_{-1}.
\label{twentyrep}
\eea
Then $({\bf 1,2})_{1}$ which has $U(1)_R$ charge 1 
describes a twistor field
$\hat{\xi}_{-\frac{1}{2},A=1}^{[BC=12]}$ 
\footnote{How one realizes the correspondence between 
twistor field and conformal supergravity field?
There is a relation given by (\ref{wab}). By substituting 
the solution of $\xi_{bC}^{[AB]}$ 
in terms of momentum space into the 
left hand side of (\ref{wab}), then there is an expression of
$\theta_a^C \int d^4 k \delta (k^2) e^{i k \cdot x} \pi_b 
\xi_{-\frac{1}{2},C}^{[AB=12]}(k)$. On the other hand, the 
right hand side of (\ref{wab}) has a term $
\int d \la^d \theta_{d}^{C} \la_a \la_b 
\hat{\xi}_{-\frac{1}{2},C}^{[AB=12]}$. Now one differentiates both 
expressions with respect to $\theta_{c}^{D}$. Then one obtains
the following relation:
$\int d^4 k \delta (k^2) e^{i k \cdot x}  
\xi_{-\frac{1}{2},C}^{[AB=12]}(k)=\int d \la^a \la_a  
\hat{\xi}_{-\frac{1}{2},C}^{[AB=12]}$. In this way, one 
gets a precise relation between 
twistor fields and spacetime fields.}.
In other words, the index $A$ plays the role of ${\bf 2}$
of $SU(2)_R$ representation. The other element of ${\bf 2}$
will be seen from the description of $g_{A=2}(Z)$ below. 
Since  $\hat{\eta}_{-\frac{1}{2}}^{B=2}$ has a representation
$({\bf 1,2})_{1}$ according to (\ref{fourrep}), 
one can construct 
the sum of these two twistor fields as 
${\cal N}=2$ twistor field,
$\hat{\widetilde{\xi}}_{-\frac{1}{2},A=1}^{[BC=12]}$
because they have common helicity and $SU(2)_R$ representation.

Under the 
$SU(4)_R \rightarrow SU(2) \times SU(2)_R \times U(1)_R$,
the $SU(4)_R$-representation ${\bf 15}$ characterized by
$V_{\mu A}^{B}$
is decomposed into 
\bea
{\bf 15} \rightarrow ({\bf 1,1})_{0} + ({\bf 1,3})_{0} +
({\bf 3,1})_{0} + ({\bf 2,2})_{2} + ({\bf 2,2})_{-2}.
\label{15rep}
\eea
Therefore,  the representation $({\bf 1,1})_{0}+({\bf 1,3})_{0}$ 
preserving
$U(1)_R$ charge corresponds to 
$\hat{V}_{-1,A}^{B}$.
We can also combine two terms in $\chi$-dependent terms
and express as 
$\hat{\widetilde{V}}_{-1,A=1}^{B=1}$.
Also from the branching rule of ${ \overline{\bf 4}}$ of $SU(4)_R$,
the representation $(\overline{\bf 1},{\bf 2})_{-1}$ gives 
a description for $ \hat{\overline{\eta}}_{-\frac{3}{2},A=1}$.
For the description of helicity on these fields, it is easy to 
check.

Let us consider when the index $A$ in $g_A(Z)$ is equal to 2.
Similarly, one can do for $g_{A=2}(Z)$:
\bea
\epsilon^{12CD}  D_{cC} D_{dD}   \; g_{A=2}(Z) 
\Big |_{\theta^3=\theta^4=0}
&=& 
\left(\la_c \la_d \right) \left[
 \left( \hat{\eta}_{-\frac{1}{2}}^{B=1} + 
\hat{\xi}_{-\frac{1}{2},A=2}^{[BC=12]} \right)+
 \psi  \left( \hat{e}_{-1} +
  \hat{V}_{-1,A=2}^{B=2} \right) \right. \nonu \\
&& \left. + 
\chi  \hat{V}_{-1,A=2}^{B=1}  +
\psi \chi  \hat{\overline{\eta}}_{-\frac{3}{2},A=2}   
\right].
\label{six}
\eea
The first two terms can be 
combined ${\cal N}=2$ twistor field
$\hat{\widetilde{\xi}}_{-\frac{1}{2},A=2}^{[BC=12]}$
because they share common helicity and $SU(2)_R$ representation 
and moreover 
the two terms in $\psi$-terms can be written as
$  \hat{\widetilde{V}}_{-1,A=2}^{B=2}$.
It is more clear if we combine the two results (\ref{five}) 
and (\ref{six}) in a single
covariant way. It becomes
$\hat{\widetilde{\xi}}_{-\frac{1}{2},A} + \psi^B  
\hat{\widetilde{V}}_{-1,A}^{B} + \psi^B \psi^C  
\hat{\overline{\eta}}_{-\frac{3}{2},A} \ep_{BC}$ which 
corresponds to the massless states $(-\frac{1}{2}, {\bf 2}), 
(-1, {\bf 3} \oplus {\bf 1}), (-\frac{3}{2}, 
{\bf 2})$ for $\pa_{\alpha} \pa_{\beta} g_{A=1,2}$ 
given in section \ref{spectrum}.
Note that 
if we use $\pa_{\alpha} \pa_{\beta}$ on $g_{A=1,2}(Z)$ 
rather than the super derivatives, then the factor $\la_c \la_d$ 
in the right hand side of (\ref{five}) or (\ref{six}) disappears. 

Now let us consider when the index $A=3$ for $g_{A}(Z)$.
In (\ref{wab}), 
one inserts the quadratic spinors $\la_a \la_b$ in front of 
${\cal N}=4$ twistor field  $g_{A=3}(Z)$ in the integrand 
which is homogeneous of
degree $-1$ and 
divide into those factors. 
Then the quantity $\la_a \la_b g_{A=3}(Z)$ is 
homogeneous of degree 1 which is the same degree of
${\cal N}=2$ twistor field we introduced in section 
\ref{spectrum}
and the quantity $\frac{d \la^c \theta^{A=3}_c }{\la_a \la_b}$
is homogeneous of degree $-1$. 
The integral appearing in (\ref{wab}) makes sense
because the whole expression is homogeneous of degree zero.
Finally, one can construct 
the following ${\cal N}=2$ object,
by acting one super derivative $D_{d,B=3}$ 
on the factor $\theta^{A=3}_{c}$
and the other super derivative $D_{d,D=4}$ 
on the $g_{A=3}(Z)$, 
\bea
\epsilon^{123D} \la_a \la_b D_{dD}   \; g_{A=3}(Z) 
\Big |_{\theta^3=\theta^4=0}
& = & 
\left( \la_a \la_b \la_{d} \right) 
\left[ \left( \hat{T}_{0[AB=34]} +   \hat{E}_{0(AB=34)}  
\right)+ \psi \left(\hat{\eta}_{-\frac{1}{2}}^{B=2}  
+ \hat{\xi}_{-\frac{1}{2},A=3}^{[BC=23]} \right) 
\right. \nonu \\
&& \left. +
\chi \left(\hat{\eta}_{-\frac{1}{2}}^{B=1}+
\hat{\xi}_{-\frac{1}{2},A=3}^{[BC=13]}
  \right) 
 +
\psi \chi  \left( \hat{e}_{-1} + \hat{V}_{-1,A=3}^{B=3}
\right) \right].
\nonu
\eea
When we act two super derivatives on $g_{A=3}(Z)$ completely, 
due to the
factor $\theta^{A=3}_{c}$ in front of it, 
the contribution will be zero after we put $\theta^3=\theta^4=0$.
If one wants to compare this twistor field with 
the spacetime field, one should insert  
the above result into the spinor integral given in (\ref{wab}).
According to the branching rule of ${ \overline{\bf 10}}$ of
$SU(4)_R$ into 
\bea
{\overline{\bf 10}} \rightarrow
({\bf 2,2})_0 + ({\bf 3,1})_{2}+ ({\bf 1,3 })_{-2}
\label{tenrep}
\eea
under the $SU(2) \times SU(2)_R \times U(1)_R$,
the twistor field $ \hat{E}_{0(AB=34)}$ describes 
one of the element for 
the representation $({\bf 3,1})_{2}$ preserving 
$U(1)_R$ charge.
From the breaking pattern (\ref{sixrep}), 
$\hat{T}_{0[AB=34]}$ which can be written as $\ep^{12CD} 
\hat{T}_{0[CD]}=\hat{T}_{0}^{[12]}$ 
corresponds to the representation 
$(\bf 1,1)_2$. 
So we combine these two terms as $
\hat{\widetilde{T}}_{0[AB=34]}$ because they share same helicity and
$SU(2)_R$ representation.
Next term can be understood from (\ref{fourrep}) and
the twistor field 
$\hat{\eta}_{-\frac{1}{2}}^{B=2}$ describes one of the element for 
the representation $({\bf 1,2})_1$.
Similarly, the twistor field $   
\hat{\xi}_{-\frac{1}{2},A=3}^{[BC=23]}$ 
represents one of the element for $({\bf 1,2})_1$ from the 
branching rule (\ref{twentyrep}). 
One can write $\hat{\widetilde{T}}_{0[AB=34]} + \psi^A  
\hat{\widetilde{\eta}}_{-\frac{1}{2},A} + \psi^B \psi^C  
\hat{\widetilde{e}}_{-1} \ep_{BC}$ 
in covariant way to which the massless states 
$(0, {\bf 1}), (-\frac{1}{2}, {\bf 2}), (-1, {\bf 1})$ given 
in section \ref{spectrum} correspond.

In addition to a chiral superfield ${\cal W}_{ab}^{AB}
$, there is also an anti chiral superfield
$\overline{\cal W}^{\dot{a}\dot{b}}_{AB}$ which depends on 
both $\hat{x}^{a\dot{a}}(= 
x^{a \dot{a}}+\theta^{aA} \overline{\theta}^{\dot{a}}_{A})$ 
and $\overline{\theta}^{\dot{a}}_A$.
Then the relation between spacetime fields and twistor fields 
can be written as
\bea
{\overline{\cal W}}^{\dot{a}\dot{b}}_{12}(\hat{x}, 
\overline{\theta}) = 
 \ep_{12AB} 
\overline{D}^{A(\dot{a}} \overline{D}^{\dot{b})B}
\int d \overline{\la}^{\dot{c}} \left[ \widetilde{g}_
{\dot{c}}(\overline{Z}) +
\hat{x}_{c\dot{c}} \widetilde{g}^{c}(\overline{Z}) + 
\overline{\theta}_{\dot{c}C} 
\widetilde{g}^C(\overline{Z}) \right]
\Big |_{\overline{\theta}_3=\overline{\theta}_4=0}
\nonu
\eea
where the complex conjugate of $Z^I$, $\overline{Z}_I$, is given by 
$\overline{Z}_I =(\overline{\la}_{\dot{a}}, 
\overline{\mu}_a = \hat{x}_{a\dot{a}} \overline{\la}^{\dot{a}},
\overline{\psi}_A = \overline{\theta}_A^{\dot{a}} 
\overline{\la}_{\dot{a}})$ using corresponding twistor equations.
The dual fields $\widetilde{g}^I(\overline{Z})$
appearing in the integrand of right hand side
have an explicit form, through Fourier-like transform,
in terms of $f^I(Z)$ \cite{BW}. 
Fourier transform maps the $f^I(Z)$ field to a dual field 
$\widetilde{g}^I(\overline{Z})$.
Inside of the integrand of this defining equation, 
there is a factor $e^{Z^K \overline{Z}_K}$.
When we differentiate $\widetilde{g}^I(\overline{Z})$ with 
respect to 
$\overline{Z}_I$, 
the factor $Z^I$ appears in the integrand, due to 
this exponential factor.
Therefore, the action of two superspace derivatives $
\overline{D}^{\dot{a}A}$'s on $\widetilde{g}^I(\overline{Z})$
leads to the two product of $Z^I$'s in the integrand, in general.
Since this integral contains an integral over 
the fermionic coordinates, the nonzero
contribution will take place only if $f^I(Z)$ does not have 
these fermionic coordinates(in the present case, 
$\theta^3$ or $\theta^4$) 
in $\theta_a^A$-expansion. In other words, we simply put 
the condition of $\theta^3=\theta^4=0$ on the $f^I(Z)$'s in order to
get ${\cal N}=2$ twistor fields from ${\cal N}=4$ field contents.  

Using similar arguments to the analysis given in $g_I(Z)$, 
one obtains
the ${\cal N}=2$ twistor field $f^I(Z)$.
From ${\cal N}=4$ twistor fields \cite{BW},
one writes
\bea
\la^a f_a(Z) 
\Big |_{\theta^3=\theta^4=0} 
 =  \hat{e}_2^{\prime} +\psi 
\hat{\overline{\eta}}_{\frac{3}{2}, A=1}^{\prime}+
\chi 
\hat{\overline{\eta}}_{\frac{3}{2}, A=2}^{\prime}+
\psi \chi \hat{\overline{T}}_1^{\prime[AB=34]}.
\label{two}
\eea
How one does understand this result?
Let us concentrate on the first term of 
(\ref{two}). 
One can see this from Fourier-like transform.
The nonzero contribution of fermionic integral will 
arise from the exponential 
factor because the first term of (\ref{two})
does not depend on the fermionic coordinate 
$\psi^A$.
Then the dependence on $\overline{\psi}_A$ of dual field
$\widetilde{g}^a(\overline{Z})$ has 
a term like $(\overline{\psi}^4)$ because
the integrals over $\psi^A$ in Fourier-like transform 
pick up $(\psi^4)$ term. 
When we go back to the expression of $g_{\dot{a}}(Z)$ 
(\ref{gadot}) and 
make a complex conjugation of it, then 
the Lorentz scalar 
$\pa_a \widetilde{g}^a(\overline{Z})$ will have a term like 
$(\overline{\psi}^4) \hat{e}_{2}^{\prime}$.
Therefore, one leads to the first term of (\ref{two}).
Other terms can be described similarly. 
Once we identify the relation between $f^I(Z)$ and 
$\widetilde{g}^I(\overline{Z})$ explicitly, 
then the spacetime interpretation 
of $f^I(Z)$ (i.e., a precise relation between hatted twistor 
fields and
unhatted spacetime fields) can be read off from those for 
$\widetilde{g}^I(\overline{Z})$ which is related to $g_I(Z)$.   

As we have seen in section \ref{spectrum},
the massless state given by $(2,{\bf 1})$ describes
$ \hat{e}_2^{\prime}$ which is a singlet of $SU(2)_R$ and 
the states $(\frac{3}{2},{\bf 2})$ corresponds to 
both $\hat{\overline{\eta}}_{\frac{3}{2}, A=1}^{\prime}$ and
$\hat{\overline{\eta}}_{\frac{3}{2}, A=2}^{\prime}$. 
The branching rule of ${ \overline{\bf 4}}$ of $SU(4)_R$ 
(\ref{fourrep}) provides
the representation $(\overline{\bf 1},{\bf 2})_{-1}$ 
with $U(1)_R$ charge $-1$ which describes $\hat{
\overline{\eta}}_{\frac{3}{2}, A}^{\prime}$. 
They are doublet ${\bf 2}$ of $SU(2)_R$ and are combined into
$ \psi^A 
\hat{\widetilde{\overline{\eta}}}_{\frac{3}{2}, A}^{\prime}$.
Next state characterized by 
$(1, {\bf 1})$ provides $\hat{\overline{T}}_1^{\prime[AB=34]}$
which is a singlet and of helicity 1.
As we have discussed before, from the branching rule 
${\bf 6}$ of $SU(4)_R$ (\ref{sixrep}), 
this corresponds to the representation  
$(\overline{\bf 1}, \overline{\bf 1})_{-2}$.
In covariant way, one can 
write it as 
$\psi_A \psi_B \hat{\overline{T}}_1^{\prime[AB]}$. 

One can analyze also the scalar function 
$\mu^{\dot{a}} f_{\dot{a}}(Z)$ where translation generators
can be diagonalized. From ${\cal N}=4$ result \cite{BW}, one requires
that other two fermionic coordinates $\theta^3$ and $\theta^4$ 
vanish. Then one gets 
\bea
\mu^{\dot{a}} f_{\dot{a}}(Z) \Big |_{\theta^3=\theta^4=0} 
& =&  \hat{e}_2 + \psi 
\hat{\overline{\eta}}_{\frac{3}{2}, A=1} +
 \chi
\hat{\overline{\eta}}_{\frac{3}{2}, A=2} +\psi \chi 
\hat{\overline{T}}_1^{[AB=34]}.  
\label{eight}
\eea
This looks like (\ref{two}) exactly except 
that the independent fields for given helicity are replaced by
other elements of `dipole' fields.
Let us consider the first term of (\ref{eight}).
The nonzero contribution of Fourier-like transform 
will arise from the exponential 
factor because the first term of (\ref{eight})
does not depend on the fermionic coordinate 
$\psi^A$.
Then the dependence on $\overline{\psi}_A$ of dual field
$\widetilde{g}_{\dot{a}}(\overline{Z})$ has 
a term like $(\overline{\psi}^4)$. 
From the expression of $g_{a}(Z)$ (\ref{ga}),  
the Lorentz scalar 
$\pa^{\dot{a}} 
\widetilde{g}_{\dot{a}}(\overline{Z})$ will have a term like 
$(\overline{\psi}^4) \hat{e}_{2}$.
Therefore, one leads to the first term of (\ref{eight}). 
The exact identification in
section \ref{spectrum} by following the analysis in previous 
paragraph can be done.
That is, the massless states 
$(2,{\bf 1}),(\frac{3}{2},{\bf 2}),(1,{\bf 1})$
given in section \ref{spectrum} 
correspond to the component twistor fields
in (\ref{eight}).

For the vertex operator with fermionic index, one can compute the
following quantity from ${\cal N}=4$ result \cite{BW}. That is, 
\bea
f^{A=1}(Z) \Big |_{\theta^3=\theta^4=0} 
& =&  \hat{\eta}_{\frac{3}{2}}^{A=1} + 
\psi ( \hat{e}_1 + \hat{V}_{1,A=1}^{B=1} ) + 
\chi  \hat{V}_{1,A=1}^{B=2}  + \psi \chi ( 
\hat{\overline{\xi}}_{\frac{1}{2}[BC=12]}^{A=1}+
\hat{\overline{\eta}}_{\frac{1}{2}, B=2}
 ).
\label{three}
\eea
One can also 
understand each term above, through Fourier-like transform.
By looking at the $(\psi^4)$ term of $g_{A=1}(Z)$ (\ref{gA}) 
and taking a 
complex conjugation with opposite helicity, 
one arrives at the first term of 
(\ref{three}).
Here a twistor field
$\hat{\overline{\xi}}_{\frac{1}{2}[BC=12]}^{A=1}$
arises from the representation $(\overline{\bf 1},{\bf 2})_{-1}$, 
according to the branching rule of ${\overline{\bf 20}}$ of
$SU(4)_R$. See for example, (\ref{twentyrep}). 
Note that 
the twistor field 
$\hat{\overline{\eta}}_{\frac{1}{2}, B=2}$ has 
same representation   
$(\overline{\bf 1},{\bf 2})_{-1}$ from the branching rule of 
${\overline{\bf 4}}$ of $SU(4)_R$.
The analysis for ${\bf 15}$ of $SU(4)_R$ representation 
(\ref{15rep}) to 
the twistor field 
$\hat{V}_{1,A=1}^{B=1}$
and $\hat{V}_{1,A=1}^{B=2}$ holds here. 
The two terms in 
$\psi$ and the two terms in $\psi \chi$ in (\ref{three}) can be  
written as
a single term $ \hat{\widetilde{V}}_
{1,A=1}^{B=1}$ and $
\hat{\overline{\widetilde{\xi}}}_{\frac{1}{2}[BC=12]}^{A=1}
$ respectively because they have same helicity and $SU(2)_R$
representation. 

Let us compute also other component 
from ${\cal N}=4$ result
\bea
f^{A=2}(Z) \Big |_{\theta^3=\theta^4=0} 
& =&  \hat{\eta}_{\frac{3}{2}}^{A=2} + 
\psi \hat{V}_{1,A=2}^{B=1}  + 
\chi ( \hat{e}_1 +  \hat{V}_{1,A=2}^{B=2} )  
+ \psi \chi ( 
\hat{\overline{\xi}}_{\frac{1}{2}[BC=12]}^{A=2} + 
\hat{\overline{\eta}}_{\frac{1}{2}, B=1} ).
\label{four}
\eea
One can write the two results (\ref{three}) and (\ref{four}) 
in $SU(2)_R$ covariant way in order to compare with the 
description given in section \ref{spectrum}.
That is, the ${\cal N}=2$ twistor field
can be summarized by 
$f^A(Z)= \hat{\eta}_{\frac{3}{2}}^{A} + \psi^B  
\hat{\widetilde{V}}_
{1,A}^{B} + \psi^B \psi^C \hat{\overline{\widetilde{\xi}}}_
{\frac{1}{2}}^{A} \ep_{BC}$. 
Each massless state coincides with $(\frac{3}{2},{\bf 2}),
(1, {\bf 3} \oplus {\bf 1}),(\frac{1}{2}, {\bf 2})$ respectively 
given in 
section \ref{spectrum}.
The helicities are encoded in the subscript of twistor fields
and $SU(2)_R$ representation 
can be read off easily.

Let us move on the final case.
One can multiply $\la_a \la_b$ in the 
integrand of Fourier-like transform
and divide $\la_a \la_b$.  
The expression $\frac{f^{A=3}(Z)}{\la_a \la_b}$
is homogeneous of
degree $-1$ which is the same degree for ${\cal N}=2$ twistor field. 
Then 
one can compute the following ${\cal N}=2$ object
which has
a super derivative  $D_{a,A=3}$, compared with
a twistor field $g_{A=3}(Z)$ on which 
a super derivative $D_{a,A=4}$ acts, 
\bea
 \ep^{12D4} \frac{1}{\la_a \la_b}  
D_{aD} f^{A=3}(Z) \Big |_{\theta^3=\theta^4=0} 
& =&   \frac{1}{\la_b } \left[ (
 \hat{e}_{1} +  \hat{V}_{1,A=3}^{B=3} ) +
\psi ( \hat{\overline{\eta}}_{\frac{1}{2}, B=1}  + 
\hat{\overline{\xi}}_{\frac{1}{2}[BC=13]}^{A=3}
) \right. \nonu \\
&&  \left. + 
\chi (  \hat{\overline{\eta}}_{\frac{1}{2}, B=2} +
\hat{\overline{\xi}}_{\frac{1}{2}[BC=23]}^{A=3} ) 
+
\psi \chi  (\hat{\overline{E}}_0^{(AB=34)} 
+ \hat{\overline{T}}_0^{[AB=34]} ) \right].
\nonu
\eea
Note that when we take the differentiation 
on $\widetilde{g}^{I=3}(\overline{Z})$ with respect to
$\overline{\theta}_4$ or $\overline{\psi}_{A=4}$, 
then a factor $\theta^4$ appears in 
the integrand of defining equation, due to the exponential factor. 
Then the fermionic integration over 
$\psi^A$ acts on $\theta^4 e^{Z^K \overline{Z}_K}  
f^{I=3}(Z)$ in the integrand.
Therefore, the nonzero contribution of this integral  
occurs when the fermionic 
integration over $\psi^{A=3}$ 
(which is equivalent to the differentiation 
$D_{a,D=3}$) acts on the function $f^{A=3}(Z)$, as above. 
Moreover, the nonzero contribution in the integrand has 
a term $\psi \chi \overline{\psi} 
\overline{\chi}$. By computing the integration over 
$\psi^{A}$, the resulting expression has a coefficient of
$\overline{\psi} \overline{\chi}$.
Therefore, the fermionic independent term of 
above equation  
$\hat{e}_{1} +  \hat{V}_{1,A=3}^{B=3}$ can be obtained
from the $\psi^A$-expansion of $g_{A=3}(Z)$ (\ref{gA}) 
with opposite 
helicity: the term which 
has a factor $\psi \chi \beta$.

The twistor field $ \hat{\overline{E}}^{0(AB=34)}$ describes 
the representation $({\bf 3},\overline{\bf 1})_{-2}$ 
in the breaking pattern 
of ${\bf 10}$ of $SU(4)_R$ which can be read off from 
(\ref{tenrep}).
By redefining the two terms in $\psi$ which have common
$SU(2)_R$ represenation and $U(1)_R$ charge due to the analysis 
of (\ref{three})  
and the two terms 
in $\chi$ which have the same 
quantum number also,  $(\overline{\bf 1},{\bf 2} )_{-1}$, 
one can write this 
as follows: 
$\hat{\overline{\widetilde{\eta}}}_{\frac{1}{2}, B=1}$ 
and 
$
\hat{\overline{\widetilde{\eta}}}_{\frac{1}{2}, B=2} 
$. 
Similar redefinition 
for the last term can be applied also by similar description given in
$g_{A=3}(Z)$ above. 
Then, we obtain the simplified form 
$ \hat{\widetilde{e}}_{1} + \psi^A \hat{\overline
{\widetilde{\eta}}}_{\frac{1}{2}, A} + \psi_A \psi_B 
 \hat{\overline{\widetilde{T}}}_0^{[AB]}$ 
which is coincident with 
the massless states $(1,{\bf 1}), (\frac{1}{2},{\bf 2}), 
(0,{\bf 1})$ given in section \ref{spectrum}.

\section{Concluding remarks}
\setcounter{equation}{0}

\indent

In this paper, 
a chiral superfield strength in 
${\cal N}=2$ conformal supergravity at linearized level was 
obtained  
by acting two superspace derivatives on ${\cal N}=4$ chiral 
superfield strength.    
By decomposing $SU(4)_R$ representation  of
${\cal N}=4$ twistor superfields into the  
$SU(2)_R$ representation of ${\cal N}=2$ twistor superfields with  
an invariant $U(1)_R$ charge, 
the physical states of 
${\cal N}=2$ conformal supergravity
can be read off from
the surviving ${\cal N}=2$ twistor superfields. 
These ${\cal N}=2$ twistor superfields
are functions of homogeneous coordinates of  
${\bf WCP}^{3|4}$.

We will describe two relevant subjects, the construction 
of ${\cal N}=3$ conformal supergravity in twistor-string theory and
a mirror symmetry of Calabi-Yau supermanifold ${\bf WCP}^{3|4}$,
in the remaining section.

Let us first 
make some comments on ${\cal N}=3$ conformal supergravity.
One can apply our analysis given in previous sections 
to ${\cal N}=3$ conformal supergravity.
The ${\cal N}=3$ chiral superfield strength \cite{Siegel,BDD,FT} 
can be 
constructed from ${\cal N}=4$ one similarly
and it is given by 
\bea
{\cal W}_{a}^{ABC}(x,\theta)= \ep^{ABC} {\cal W}_a (x, \theta)=
\ep^{ABCD} D_{aD} {\cal
W}^{{\cal N}=4}(x,\theta) \Big |_{\theta^4=0}
\nonu
\eea 
explicitly.   
Then we expect to have the following $\theta$-expansion:
${\cal W}_{a}^{123}(x,\theta)=\La_{4a} +
\cdots + (\theta)^6 \pa^{\mu} 
\pa_{\mu} \pa_{a\dot{a}} \overline{\La}^{4\dot{a}}$.
The ${\cal N}=3$ Weyl multiplet is a Lorentz spinor with 
$\La_{4a}$ as its lowest $\theta$ component $\La_{4a}$.
This can be done by computing the  
above relation, as we have done for ${\cal N}=2$ conformal 
supergravity. 
The field contents have extra spinor field $\La_A^a$ and complex
scalar $E_{(AB)}$ as well as 
${\cal N}=2$ field contents with different multiplicities
we have discussed so far.
The constraint equation for ${\cal N}=3$ 
conformal supergravity can be read off from the 
${\cal N}=4$ relation (\ref{constraint}).
That is,
\bea
 \ep^{ABC} D_{Da} D_{E}^b D^{a}_A 
{\cal W}_b = \ep_{DEF} \overline{D}^{B \dot{a}} 
\overline{D}^{C}_{\dot{b}} 
\overline{D}_{\dot{a}}^F 
\overline{\cal W}^{\dot{b}}. 
\nonu
\eea

Since the square of the Weyl multiplet, contracted over
all indices, is a chiral scalar multiplet with Weyl weight
$w=1$. The highest component, a Lorentz scalar 
has Weyl weight $w=4$. 
The action at the linearized level is given by
\bea
S= \int d^4 x \int d^6 \theta \; {\cal W}^2, \qquad
{\cal W}^2 = \ep_{ABC} \ep_{DEF} {\cal W}^{aABC}
{\cal W}_{a}^{DEF}.  
\nonu
\eea
By computing the $\theta$-integrals, 
the bosonic action is obtained. The extra piece consists of 
$\Lambda_a \pa^{\mu} \pa_{\mu} 
\pa^{a \dot{a}} \overline{\Lambda}_{\dot{a}}$
and $E_A \pa_{\mu} \pa^{\mu} \overline{E}^A$.
Then the physical spectrum of conformal supergravity 
for these extra piece can be obtained by solving the equations 
of motion. As in ${\cal N}=4$ conformal supergravity, a spinor
field $\Lambda_a$ has a helicity $-1/2,-1/2,1/2$ while 
complex scalar field $E_A$ has zero helicity.

Using $SU(3)_R$-invariant epsilon tensor $\ep^{ABC}$, one can 
write ${\cal N}=3$ conformal supergravity field contents 
\cite{FT} 
from the field contents of
${\cal N}=4$ conformal supergravity as follows:
\bea
&& \ep^{ABCD} \La_D = \ep^{ABC} \La, \;\;\; 
\ep^{ABCD} E_{DE}=\ep^{ABC} E_E, \;\;\;
\ep^{ABC} T_{(ab)C}, \;\;\; \ep^{ABD} \xi_{aCD}, \nonu \\
&& \eta_{\mu}^{Aa},
\;\;\; V_{\mu A}^{B},
\;\;\; A_{\mu}, \;\;\; \ep^{ABC} \ep_{DEF} d^{F}_{C}, 
\;\;\; e_{\mu}^{a\dot{a}}, 
\;\;\; \overline{\eta}_{\mu A}^{\dot{a}}, 
  \;\;\; \ep_{ABD} \overline{\xi}^{\dot{a} CD}, \nonu \\
&&
\ep_{ABC} \overline{T}_{(\dot{a} \dot{b})}^C, \;\;\;
\ep_{ABCE} \overline{E}^{DE} =
\ep_{ABC} \overline{E}^D, \;\;\; \ep_{ABCD}
\overline{\La}^D = 
\ep_{ABC} \overline{\La}
\nonu
\eea
where the ${\cal N}=4$ fields $ \La^{ABC} \equiv 
\ep^{ABCD} \La_D$ and
$E^{ABC}_{E} \equiv \ep^{ABCD} E_{DE}$ 
are written as in dual form 
which will make the truncation of ${\cal N}=4$ conformal supergravity
more clear.  
Under the breakings of $SU(4)_R$ representations of ${\cal N}=4$
theory
into $SU(3)_R 
\times U(1)_R$, 
\bea
{\overline{\bf 4}} \rightarrow 
\overline{{\bf 3}}_{-1} +
{\bf 1}_{3}, \qquad 
\overline{\bf 10} 
\rightarrow {\bf 1}_{6} +\overline{{\bf 3}}_{2} +
\overline{{\bf 6}}_{-2}, \qquad 
{\bf 6} \rightarrow {\bf 3}_{-2} +
\overline{\bf 3}_{2}
\nonu
\eea
the above ${\cal N}=3$ conformal supergravity 
$\La, E_E, T_{(ab)C}$ describe
${\bf 1}_{3}, \overline{{\bf 3}}_{2}, \overline{\bf 3}_{2}$ 
by realizing the correct $U(1)_R$ charge respectively \footnote{
Similarly, for  the conjugated representations 
$\overline{\La}, \overline{E}^D, 
\overline{T}_{(\dot{a} \dot{b})}^{C}$, one can identify  
${\bf 1}_{-3}, {\bf 3}_{-2}, {\bf 3}_{-2}$ with them respectively.  
For the breaking patterns
$
{\bf 20} \rightarrow  {\bf 3}_{1} + 
\overline{\bf 3}_{5} +
\overline{\bf 6}_{1} + {\bf 8}_{-3}$,
${{\bf 4}} \rightarrow 
{{\bf 3}}_{1} +
{\bf 1}_{-3}$, 
$
{\bf 15} \rightarrow {\bf 1}_{0} + 
{\bf 3}_{4} +
\overline{\bf 3}_{-4} +{\bf 8}_{0}$
the above conformal supergravity fields
$\xi_{aCD}, \eta_{\mu}^{Aa}, V_{\mu A}^{B}, A_{\mu}$
correspond to 
$ {\bf 3}_{1}+\overline{\bf 6}_{1}, 
{\bf 3}_{1}, {\bf 8}_{0}, {\bf 1}_0$ respectively.
For $\overline{\eta}_{\mu A}^{\dot{a}}$ 
and $\overline{\xi}^{aCD}$,
one identifies them with $\overline{{\bf 3}}_{-1}$ and $
 \overline{{\bf 3}}_{-1}+{\bf 6}_{-1}$ respectively.
Finally, under the breaking pattern 
$
{\bf
  20'} \rightarrow \overline{{\bf 6}}_{4} +{\bf 6}_{-4} +
{\bf 8}_{0}$, 
the field $ d^{F}_{C}$ describes a representation 
${\bf 8}_{0}$.}.
A zero total number of on-shell degree of freedom 
can be checked \footnote{For the extra fields, 
it is known that the number of on-shell degrees of 
freedom is given by $\nu(E_E)=\nu(\overline{E}^D)=1$ and 
$ \nu(\La) =
\nu(\overline{\La}) =-3$ \cite{FT} which can be seen also 
from the helicity counting above.} 
by multiplying each degree of freedom 
by the number of each type of the field in the spectrum
$
\nu_{\mbox{total}} =+48-48=0$.

The structure of the off-shell multiplets  
of ${\cal N}=3$ conformal supergravity is in correspondence
with the structure of the on-shell massive supermultiplet
of extended supersymmetry \cite{BDD}.
This can be done by establishing a correspondence between 
the states of each particular spin. 
The massive supermultiplets are described by 
an antisymmetric tensor representation of $Sp(6)$  and 
by decomposing these representations into the $SU(3)_R$ subgroup,
the number of states for given spins are determined.
For example, the number of spin 1 states can be described by
${\bf 15} \rightarrow {\bf 1} +{\bf 8} +{\bf 3} +
\overline{\bf 3}$ under $Sp(6) \rightarrow SU(3)_R$.
This is in agreement with the spectrum of conformal
supergravity: 1 $U(1)$ and 8 $SU(3)_R$ vectors and 3 antiself-dual
and 3 self-dual antisymmetric tensors. 

For ${\cal N}=3$ conformal supergravity, at least 
there should be present three fermionic coordinates of
weight 1 which will play a role for three fermionic coordinates 
in ${\cal N}=3$ superspace. 
Then the only possible Calabi-Yau 
supermanifold for four dimensional fermionic submanifold
is nothing but ${\bf CP}^{3|4}$. 
The action of ${\cal N}=3$ superconformal algebra
 $SU(2,2|3)$ on the homogeneous coordinates
$(\la^a, \mu^{\dot{a}}, \psi, \chi, \alpha)$
of ${\bf CP}^{3|4}$ can be generated 
by $7 \times 7$ supertraceless matrices where
the trace of $4 \times 4$ upper-left corner is
equal to the trace of $3 \times 3$ lower-right corner.
By checking the breaking patterns $SU(4)_R \rightarrow SU(3)_R 
\times U(1)_R$, and reading off the correct $U(1)_R$ charge,
one gets the full spectrum of massless fields. 
Similar but different symmetry breaking pattern 
$SU(4)_R \rightarrow SU(3) \times U(1)_R$ preserving only
${\cal N}=1$ supersymmetry appears in 
the context of  
marginal deformations of ${\cal N}=4$ super Yang-Mills
theory from open/closed  twistor-string theory \cite{KZ}.

The helicity states by the ${\cal N}=3$ twistor fields
$f^{I}(Z)$ can be summarized as follows:
\bea
\la^a f_a &:& (2, {\bf 1}), 
\;\; (\frac{3}{2}, \overline{\bf 3}), \;\;
(1, {\bf 3}), \;\; (\frac{1}{2}, {\bf 1}), \nonu \\
\mu^{\dot{a}} f_{\dot{a}} &:& (2, {\bf 1}), \;\; (\frac{3}{2}, 
\overline{\bf 3}), 
\;\; (1, {\bf 3}), \;\; (\frac{1}{2}, {\bf 1}), \nonu \\
f^{A=1,2,3} &:& (\frac{3}{2}, {\bf 3}), \;\; 
(1, {\bf 8} \oplus {\bf 1}), \;\; (\frac{1}{2}, 
\overline{\bf 3}  +{\bf 6} ), 
\;\; (0, {\bf 3}), 
\nonu \\
\pa_{\beta} f^{A=4} &:& (1, {\bf 1}), \;\; 
(\frac{1}{2}, \overline{\bf 3}), 
\;\; (0, {\bf 3}), \;\; (-\frac{1}{2}, {\bf 1} ). 
\nonu
\eea
In the last column elements which do not exist in ${\cal N}=2$
conformal supergravity we have discussed so far, 
one can see  the presence of 
twistor spinor fields 
$\hat{\overline{\La}}^{\prime A=4}_{\frac{1}{2}}, 
\hat{\overline{\La}}^{A=4}_{\frac{1}{2}}$, 
three complex scalars $\hat{\overline{E}}_0^{(A, B=4)}$, and
other kind of spinor field 
$\hat{\overline{\La}}^{A=4}_{-\frac{1}{2}}$ for 
the appropriate helicity and $SU(3)_R$ representation.
Each row has 
bosonic degree of freedom 4 and fermionic degree of freedom $-4$.
There exists six columns. Therefore, there are 24 bosonic 
and $-24$ bosonic degrees of freedom. 
It is rather straightforward to determine twistor fields 
from ${\cal N}=4$ ones explicitly 
\footnote{For example, let us compute $\la^a f_a(Z) 
\Big |_{\theta^4=0}$. By putting the condition of $\theta^4=0$ into
the ${\cal N}=4$ expression $\la^a f_a(Z)$, one gets  
$\hat{e}_2^{\prime} + \psi^A 
\hat{\overline{\eta}}_{\frac{3}{2}, A}^{\prime} 
+\psi^A \psi^B \ep_{ABC}
\hat{\overline{T}}_1^{[C,4]}  + \psi^A \psi^B \psi^C
\ep_{ABC}
\overline{\Lambda}_{\frac{1}{2}}^{\prime D=4}$ where $\psi^A$
transforms as ${\bf 3}$ under the $SU(3)_R$.
That is, $\psi^1=\psi, 
\psi^2 = \chi, \psi^3 =\alpha$. 
Let us concentrate on the first term. 
One can see this from Fourier-like transform.
The nonzero contribution of fermionic integral will 
arise from the exponential 
factor.
Then the dependence on $\overline{\psi}_A$ of dual field
$\widetilde{g}^a(\overline{Z})$ has 
a term like $(\overline{\psi}^4) \hat{e}_2^{\prime}$ because
the integrals over $\psi^A$ in Fourier-like transform 
pick up $(\psi^4)$ term. 
What about the top component? That comes from 
the component field with a  
term like $(\overline{\psi}_{A=4})$
in dual field
$\widetilde{g}^a(\overline{Z})$ because the fermionic integral gives
a nonzero result for a quartic term in $\psi^A$. Three of them 
arises from $f_a(Z)$ above and one of them from the exponential 
factor.}. 
 
Similarly, 
the helicity states by the ${\cal N}=3$ twistor fields
$g_{I}(Z)$ can be summarized by 
\bea
\pa_{\beta} \; \pa_a g^a &:& (-\frac{1}{2}, {\bf 1}), 
\;\; (-1, \overline{\bf 3}), \;\;
(-\frac{3}{2}, {\bf 3}), \;\; (-2, {\bf 1}), \nonu \\
\pa_{\beta} \; \pa_{\dot{a}} g^{\dot{a}} &:& (-\frac{1}{2}, 
{\bf 1}), \;\; (-1, 
\overline{\bf 3}), 
\;\; (-\frac{3}{2}, {\bf 3}), \;\; (-2, {\bf 1}), \nonu \\
\pa_{\beta} \; g_{A=1,2,3} &:& (0, \overline{{\bf 3}}), \;\; 
(-\frac{1}{2}, 
{\bf 3}  +\overline{{\bf 6}} ), \;\;
(-1, {\bf 8} \oplus {\bf 1}), 
\;\; (-\frac{3}{2}, \overline{\bf 3}), 
\nonu \\
g_{A=4} &:& (\frac{1}{2}, {\bf 1}), \;\; 
(0, \overline{\bf 3}), 
\;\; (-\frac{1}{2}, {\bf 3}), \;\; (-1, {\bf 1} ). 
\nonu
\eea
It is evident from the corresponding (\ref{wab}) with a single
$D_{4a}$ that by chain rule one can calculate the objects 
acting on $g_b(Z), g^{\dot{b}}(Z),$ and $\theta^A_b g_A(Z)$.
When we compute the last one, there is no contribution from 
$\theta^{A=4}_b D_{4a} \;
g_{A=4}(Z)$ because at the final expression 
we put $\theta^4=0$. This is the reason why there is a 
super derivative acting on $g_{A=1,2,3}(Z)$ except 
$g_{A=4}(Z)$ 
above helicity states.  
In this case, 
the 
twistor spinor fields 
$\hat{{\La}}_{-\frac{1}{2},A=4}, 
\hat{{\La}}^{\prime}_{-\frac{1}{2},A=4}$, 
three complex scalars $\hat{E}_{(0,A, B=4)}$, and
other kind of spinor field 
$\hat{\La}^{A=4}_{\frac{1}{2}, A=4}$ for 
the appropriate helicity and $SU(3)_R$ representation
appear in the first columns.

There are also 24 bosonic and $-24$ fermionic
degrees of freedom. 
A zero total number of on-shell(dynamical) degree of freedom 
is given by 48 bosonic and $-48$ fermionic degrees of freedom, 
by counting each bosonic and fermionic degree of
freedom 
in the full spectrum.

We will consider the second subject.
The  weighted projective superspace
${\bf WCP }^{3|4}$ has 
an extension of a linear sigma model 
description in terms of four bosonic homogeneous 
coordinates of weight 1 and four fermionic 
homogeneous coordinates of weights 1, 1, $-1$, and 3, 
respectively.
The geometry of the linear sigma model 
with a given complexified Kahler class 
parameter  can be analyzed by 
solving the D-term constraint.
The super Landau Ginzburg B-model mirror of  
this Calabi-Yau supermanifold is
given by the path integral for the holomorphic sector
\bea
\int  \prod_{I=1}^{4} d Y_I  
\prod_{J=1}^{4} d X_J d \eta_J \d \chi_J 
\de \left(\sum_{I=1}^{4} Y_I- X_1-X_2 + X_3 -3X_4 -t  \right)  
\nonu \\
 \times \exp \left[\sum_{I=1}^{4} e^{-Y_I}  + \sum_{J=1}^{4} 
e^{-X_J} \left(1+ \eta_J \chi_J \right) \right].
\nonu
\eea
The super 
Landau Ginzburg model has 
7 bosonic(1 degree of freedom is removed by
delta function constraint) and 8 fermionic degrees of 
freedom. The mirror manifold has the same super dimension 
$7-8=-1$ as the original one $3-4=-1$ by mirror symmetry.

With the same spirit of \cite{AV,Ahnjuly}, 
it would be interesting to take 
A-model on  ${\bf WCP}^{3|4}$ which is mapped 
to B-model on the same space by S-duality and study 
the B-model mirror of the topological A-model on
${\bf WCP}^{3|4}$.
It would be also interesting to see 
where they lead to some
hypersurface equation satisfied by mirror 
Calabi-Yau supermanifold and 
its mirror geometry.  

Recently, the topological B-model on fattened complex 
manifolds \cite{Sae},
which are an extensions of ordinary manifolds with additional 
dimensions by even nilpotent  coordinates, has been studied. 
It would be interesting to see this kind of  
exotic supermanifolds
corresponding to ${\bf WCP}^{3|4}$.
A map between a Kahler spacetime foam in conformal 
supergravity and twistor space carrying D1 brane charge 
was found \cite{HP}. It is also interesting to 
see how this mapping arises in our construction. 
As an open subset of the weighted projective space, the so-called
enhanced super twistor space \cite{Wolf} has been found in the 
construction of extended self-dual super Yang-Mills hierarchies.
It is also interesting to see how they appear in ${\cal N} \leq 4$ 
conformal
supergravity sector.   

\vspace{1cm}
\centerline{\bf Acknowledgments}
\indent

I would like to thank N. Berkovits, A.D. Popov, P. Svrcek and 
E. Witten for discussions. 
This research was supported by a grant in aid from the 
Monell Foundation through Institute for Advanced Study, 
by SBS Foundation, 
and by Korea Research Foundation Grant (KRF-2002-015-CS0006).


\begin{thebibliography}{99}

\bibitem{Witten}
E.~Witten,
``Perturbative gauge theory as a string theory in twistor space,''
Commun.\ Math.\ Phys.\  {\bf 252}, 189 (2004)
[arXiv:hep-th/0312171].

\bibitem{Berk}
N.~Berkovits,
``An alternative string theory 
in twistor space for N = 4 super-Yang-Mills,''
Phys.\ Rev.\ Lett.\  {\bf 93}, 011601 (2004)
[arXiv:hep-th/0402045].

\bibitem{Pen}
R.~Penrose,
``Twistor Quantization and Curved Space-Time,''
Int.\ J.\ Theor.\ Phys.\ {\bf 1} (1968) 61.

\bibitem{Ati}
M.~F.~Atiyah,
``Geometry of Yang-Mills Fields,''
Lezioni Fermiane (Academia Nazionale dei Lincei Scuola Normale
Superiore), Pisa, 1979.

\bibitem{BW}
N.~Berkovits and E.~Witten,
``Conformal supergravity in twistor-string theory,''
JHEP {\bf 0408}, 009 (2004)
[arXiv:hep-th/0406051].

\bibitem{Siegel}
W.~Siegel,
``On-Shell O(N) Supergravity In Superspace,''
Nucl.\ Phys.\ B {\bf 177}, 325 (1981).

\bibitem{BDD}
E.~Bergshoeff, M.~de Roo and B.~de Wit,
``Extended Conformal Supergravity,''
Nucl.\ Phys.\ B {\bf 182}, 173 (1981).

\bibitem{FT}
E.~S.~Fradkin and A.~A.~Tseytlin,
``Conformal Supergravity,''
Phys.\ Rept.\  {\bf 119}, 233 (1985).

\bibitem{FZ}
S.~Ferrara and B.~Zumino,
``Structure Of Linearized Supergravity and Conformal Supergravity,''
Nucl.\ Phys.\ B {\bf 134}, 301 (1978).

\bibitem{Ahnsept}
C.~Ahn,
``N = 1 conformal supergravity and twistor-string theory,''
JHEP {\bf 0410}, 064 (2004)
[arXiv:hep-th/0409195].

\bibitem{PW}
A.~D.~Popov and M.~Wolf,
``Topological B-model on weighted projective spaces and 
self-dual models in
four dimensions,''
JHEP {\bf 0409}, 007 (2004)
[arXiv:hep-th/0406224].

\bibitem{Sae}
C.~Saemann,
``The topological B-model 
on fattened complex manifolds and subsectors of N =
4 self-dual Yang-Mills theory,''
[arXiv:hep-th/0410292].

\bibitem{gs}
W.~Siegel and S.~J.~Gates,
``Superprojectors,''
Nucl.\ Phys.\ B {\bf 189}, 295 (1981).

\bibitem{superspace}
S.~J.~Gates, M.~T.~Grisaru, M.~Rocek and W.~Siegel,
``Superspace, Or One Thousand And One Lessons In Supersymmetry,''
Front.\ Phys.\  {\bf 58}, 1 (1983)
[arXiv:hep-th/0108200].

\bibitem{WB}
J.~Wess and J.~Bagger,
``Supersymmetry and supergravity,'' Princeton University 
Press, 1992

\bibitem{Slansky}
R.~Slansky,
``Group Theory For Unified Model Building,''
Phys.\ Rept.\  {\bf 79}, 1 (1981).

\bibitem{KZ}
M.~Kulaxizi and K.~Zoubos,
``Marginal deformations of N = 4 SYM 
from open / closed twistor strings,''
[arXiv:hep-th/0410122].

\bibitem{AV}
M.~Aganagic ~and C.~Vafa,
~``Mirror symmetry and supermanifolds,''
[arXiv:hep-th/0403192].

\bibitem{Ahnjuly}
C.~Ahn,
``Mirror symmetry of Calabi-Yau supermanifolds,''
[arXiv:hep-th/0407009].

\bibitem{HP}
S.~A.~Hartnoll and G.~Policastro,
``Spacetime foam in twistor string theory,''
[arXiv:hep-th/0412044].

\bibitem{Wolf}
M.~Wolf,
``On Hidden Symmetries 
of a Super Gauge Theory and Twistor String Theory,''
[arXiv:hep-th/0412163].



\end{thebibliography}
\end{document}